\documentclass[reprint, showpacs, superscriptaddress, aps, twocolumn, pre, 10pt]{revtex4-1}
\usepackage{graphicx}
\usepackage{dcolumn}
\usepackage{color}
\usepackage{amssymb}
\usepackage{amsmath}
\usepackage{bm}
\usepackage{textcomp}
\usepackage{tabularx}
\usepackage{hyperref}
\usepackage{natbib} 
\usepackage[usenames,dvipsnames]{xcolor}
\hypersetup{
colorlinks,
citecolor=blue,
filecolor=blue,
linkcolor=blue,
urlcolor=blue}

\begin{document}
\title{Similarities between 2D and 3D convection for large Prandtl number}

\author{Ambrish Pandey}
\email{pambrish@iitk.ac.in}
\author{Mahendra K. Verma}
\email{mkv@iitk.ac.in}
\author{Anando G. Chatterjee}
\author{Biplab Dutta}
\affiliation{Department of Physics, Indian Institute of Technology, Kanpur 208016, India}
\date{\today}

\begin{abstract}
Using direct numerical simulations of Rayleigh-B\'{e}nard convection (RBC),  we perform a comparative study of  the spectra and fluxes of energy and entropy, and the scaling of large-scale quantities for large and infinite Prandtl numbers in two (2D) and three (3D) dimensions.  We observe close similarities between the 2D and 3D RBC, in particular the kinetic energy spectrum $E_u(k) \sim k^{-13/3}$, and the entropy spectrum exhibits a dual branch with a dominant $k^{-2}$ spectrum.  We showed that  the dominant Fourier modes in the 2D and 3D flows are very close.  Consequently, the 3D RBC is quasi two-dimensional, which is the reason for the similarities between the 2D and 3D RBC for large- and infinite Prandtl numbers.
\end{abstract}


\maketitle
\section{Introduction}\label{sec:intro}

Thermal convection is an important mode of heat transport in the interiors of stars and planets, as well as in many engineering applications.  Rayleigh-B\'{e}nard convection (RBC) is an idealized model of thermal convection, in which a fluid, placed between two horizontal thermally conducting plates, is heated from the bottom and cooled from the top~\cite{Ahlers:RMP2009}. The resulting convective motion is primarily governed by two nondimensional parameters, the Rayleigh number $\mathrm{Ra}$, which  is the ratio between the buoyancy and viscous force, and the Prandtl number $\mathrm{Pr}$, which is the ratio between the kinematic viscosity and thermal diffusivity. 

Earth's mantle and viscous fluids have large Prandtl number, and their convective flow is dominated by sharp ``plumes". Schmalzl \textit{et al.}~\cite{Schmalzl:GAFD2002, Schmalzl:EPL2004} and van der Poel \textit{et al.}~\cite{Poel:JFM2013} showed that for large Prandtl number, the flow structures and global quantities, e.g., the Nusselt number and Reynolds number, exhibit similar behaviour for three dimensions (3D) and two dimensions (2D).    In the present paper, we analyze the flow behavior of 2D and 3D flows for large Prandtl numbers, and show that the flow in the third direction in 3D RBC gets suppressed, and the large-scale Fourier modes of 2D and 3D RBC are very similar.

The energy and entropy spectra are important quantities in Rayleigh-B\'{e}nard convection, and have been studied extensively for various Prandtl numbers~\cite{Lohse:ARFM2010, Grossmann:PRA1992, Lvov:PRL1991, Lvov:PD1992, Toh:PRL1994, Vincent:PRE1999, Vincent:PRE2000, Mishra:PRE2010, Pandey:PRE2014}. Pandey \textit{et al.}~\cite{Pandey:PRE2014}, in their numerical simulations for very large Prandtl numbers in three dimensions, reported that the kinetic energy spectrum $E_u(k)$ scales as $k^{-13/3}$, and the entropy spectrum $E_\theta(k)$ shows a dual branch with a dominant $k^{-2}$ spectrum.  They also showed that the scaling of the energy and entropy spectra are similar for the free-slip and no-slip boundary conditions, apart from the prefactors. 

In this paper, we performed 2D and 3D RBC simulations  for the Prandtl numbers $10^2, 10^3$, and $\infty$, and the Rayleigh numbers between $10^5$ and $5 \times 10^8$.  We compute the ten most dominant Fourier modes of 2D and 3D flows, and show them to be very close, which is the reason for the similarities between 2D and 3D RBC.  We compute the spectra and fluxes of energy and entropy for 2D and 3D flows, and show them to be very similar.  We also show that the viscous and thermal dissipation rates for 2D and 3D RBC behave similarly.  For completeness and validation, we demonstrate similarities between the Nusselt and P\'{e}clet numbers and temperature fluctuations for 2D and 3D RBC, consistent with the earlier results of  Schmalzl \textit{et al.}~\cite{Schmalzl:GAFD2002, Schmalzl:EPL2004}, van der Poel \textit{et al.}~\cite{Poel:JFM2013}, and Silano \textit{et al.}~\cite{Silano:JFM2010}.

The paper is organized as follows: In Sec.~\ref{sec:equations}, we discuss the governing equations for large- and infinite Prandtl numbers. Details of our numerical simulations are provided in Sec.~\ref{sec:numerical}. In Sec.~\ref{sec:similarity}, we compare the most dominant Fourier modes of 2D and 3D RBC for $\mathrm{Pr}=\infty$.  In Sec.~\ref{sec:spectra}, we discuss the spectra and fluxes of the kinetic energy and entropy. Scaling of large-scale quantities such as the Nusselt and P\'{e}clet numbers, the temperature fluctuations, and the viscous and thermal dissipation rates are discussed in Sec.~\ref{sec:results}. We conclude in Sec.~\ref{sec:conclusion}. 

\section{Governing equations}\label{sec:equations}
The equations of Rayleigh-B\'{e}nard convection under Boussinesq approximation for a fluid confined between two plates separated by a distance $d$ are 
\begin{eqnarray}
\frac{\partial {\bf u}}{\partial t} + ({\bf u} \cdot \nabla){\bf u} & = & - \nabla \sigma + \theta \hat{z} + \sqrt{\frac{\mathrm{Pr}}{\mathrm{Ra}}} \nabla^2 {\bf u}, \label{eq:u_non} \\
\frac{\partial \theta}{\partial t} + ({\bf u} \cdot \nabla)\theta & = & u_z + \frac{1}{\sqrt{\mathrm{Pr Ra}}} \nabla^2 \theta, \label{eq:th_non} \\
\nabla \cdot {\bf u} & = & 0, \label{eq:cont}
\end{eqnarray}
where ${\bf u}=(u_x, u_y, u_z)$ is the velocity field, $\theta$ and $\sigma$ are the deviations of the temperature and pressure fields from the conduction state, and $\hat{z}$ is the buoyancy direction. The two nondimensional parameters are Rayleigh number $\mathrm{Ra} = \alpha g \Delta d^3 /\nu \kappa$ and the Prandtl number $\mathrm{Pr} = \nu/\kappa$, where $\Delta$ is the temperature difference between top and bottom plates, $g$ is the acceleration due to gravity, and $\alpha$, $\nu$, and $\kappa$ are the heat expansion coefficient, kinematic viscosity, and thermal diffusivity of the fluid, respectively. The above nondimensional equations are obtained by using $d$, $\sqrt{\alpha g \Delta d}$, and $\Delta$ as the length, velocity, and temperature scales, respectively.

For very large Prandtl number, $\sqrt{\alpha g \Delta d / \mathrm{Pr}}$ is used as the velocity scale for the nondimensionalization, which yields   
\begin{eqnarray}
\frac{1}{\mathrm{Pr}}\left[ \frac{\partial {\bf u}}{\partial t} + ({\bf u} \cdot \nabla){\bf u} \right] & = & - \nabla \sigma + \theta \hat{z} + \frac{1}{\sqrt{\mathrm{Ra}}} \nabla^2 {\bf u}, \label{eq:u_non_inf} \\
\frac{\partial \theta}{\partial t} + ({\bf u} \cdot \nabla)\theta & = & u_z + \frac{1}{\sqrt{\mathrm{Ra}}} \nabla^2 \theta, \label{eq:th_non_inf} \\
\nabla \cdot {\bf u} & = & 0. \label{eq:cont_inf}
\end{eqnarray}
In the limit of infinite Prandtl number, Eq.~(\ref{eq:u_non_inf}) reduces to a linear equation~\cite{Pandey:PRE2014}
\begin{equation}
- \nabla \sigma + \theta \hat{z} + \frac{1}{\sqrt{\mathrm{Ra}}} \nabla^2 {\bf u} = 0. \label{eq:u_inf}
\end{equation}
In the Fourier space, the above equation transforms to
\begin{eqnarray}
-i {\mathbf k} \hat{\sigma}(\mathbf k) + \hat{\theta}(\mathbf k) \hat{z}- \frac{1}{\sqrt{\mathrm{Ra}}} k^2 \hat{\mathbf u}(\mathbf k) & = & 0, \label{eq:u_k}
\end{eqnarray}
where $\hat{\sigma}$, $\hat{\theta}$, and $\hat{{\bf u}}$, are the Fourier transforms of $\sigma$, $\theta$, and ${\bf u}$, respectively, and ${\bf k}=(k_x, k_y, k_z)$ is the wave vector. Using the constraint that the flow is divergence-free, i.e., ${\mathbf k} \cdot  \hat{\mathbf u}({\mathbf k}) = 0$, the velocity and pressure fields can be expressed in terms of temperature fluctuations as~\cite{Pandey:PRE2014}
\begin{eqnarray}
 \hat{\sigma}(\mathbf k) & = & -i \frac{k_z}{k^2} \hat{\theta}(\mathbf k), \label{eq:p_k} \\
\hat{u}_z (\mathbf k) & = & \sqrt{\mathrm{Ra}}  \frac{k_\perp^2}{k^4} \hat{\theta}(\mathbf k), \label{eq:u_z} \\
\hat{u}_{x,y} (\mathbf k) & = & - \sqrt{\mathrm{Ra}}  \frac{k_z k_{x,y}}{k^4} \hat{\theta}(\mathbf k), \label{eq:u_xy}
\end{eqnarray}
where $k_\perp^2 = k_x^2 + k_y^2$ in 3D, and $k_\perp^2 = k_x^2$ in 2D (assuming $k_y=0$). Using these relations, the kinetic energy $E_u$ can be expressed in terms of entropy as 
\begin{equation}
E_u({\bf k}) = \frac{1}{2} |\hat{{\mathbf u}} ({\mathbf k})|^2 = \frac{1}{2} \mathrm{Ra} \frac{k_\perp^2}{k^6} |\hat{\theta} ({\mathbf k})|^2 = \mathrm{Ra} \frac{ k_\perp^2}{k^6} E_\theta({\bf k}). \label{eq:Eu_Eth}
\end{equation}

For the $\mathrm{Pr} = \infty$ limit, the nonlinear term for the velocity field, $({\bf u} \cdot \nabla){\bf u}$ is absent, 
and  the pressure, buoyancy, and  viscous terms are comparable to each other.  Assuming that the large-scale Fourier modes dominate the flow, we can estimate the ratios of these terms by computing them for the most dominant  $\mathrm u(\mathbf k)$ that occurs for ${\mathbf k} = (\pi/\sqrt{2},0,\pi)$. Hence, the aforementioned ratios can be estimated to be approximately
\begin{eqnarray}
\frac{|\theta|}{|\nabla \sigma|} \approx \frac{|\theta({\bf k}) |}{|k \sigma({\bf k})|} & \approx & \frac{k}{k_z} \approx 1  \label{eq:buoyancy_pressure}, \\
\frac{|\theta|}{|\nabla^2 {\mathbf u}|/\sqrt{\mathrm{Ra}} } \approx \frac{|\theta({\bf k}) |}{|k^2 {\mathbf u}({\bf k})/\sqrt{\mathrm{Ra}}|} & \approx & \frac{k}{k_\perp} \approx 1.  \label{eq:theta_visc} 
\end{eqnarray}

For very large $\mathrm{Pr}$, the nonlinear term for the velocity field, $({\bf u} \cdot \nabla){\bf u}$ is weak, consequently the kinetic energy flux is very weak in this regime.   The flow is dominated by  the  pressure, buoyancy, and viscous terms similar to that for the $\mathrm{Pr} = \infty$ limit. The nonlinearity of the temperature equation, $({\bf u} \cdot \nabla)\theta$, however is quite strong, and it yields a finite entropy flux for large- and infinite Prandtl numbers. We will demonstrate this statement using numerical data.  
 
In this paper, we solve RBC for  large- and infinite $\mathrm{Pr}$;  for large $\mathrm{Pr}$, we solve Eqs.~(\ref{eq:u_non_inf}$-$\ref{eq:cont_inf}), while for $\mathrm{Pr}=\infty$, we solve Eqs.~(\ref{eq:u_inf}, \ref{eq:th_non_inf}, \ref{eq:cont_inf}).  In the next section, we describe the numerical method used for our simulations.

\section{Numerical Method} \label{sec:numerical}
We solve the governing equations [Eqs.~(\ref{eq:u_non_inf}$-$\ref{eq:cont_inf})] for large Prandtl numbers, and Eqs.~(\ref{eq:u_inf}, \ref{eq:th_non_inf}, \ref{eq:cont_inf}) for $\mathrm{Pr} = \infty$.  The box geometry of the 2D simulations is $2\sqrt{2}:1$, and that for the 3D simulations is $2\sqrt{2}:2\sqrt{2}:1$.  For the horizontal plates, we employ stress-free boundary condition for the velocity field, and conducting boundary condition for the temperature field.  However, for the vertical side walls,  periodic boundary condition is used for both the temperature and velocity fields.  The fourth order Runge-Kutta method is used for the time advancement, and 2/3 rule for dealiasing. We use  the pseudospectral code TARANG~\cite{Verma:Pramana2013} for our simulations.  More details about the numerical scheme can be found in Ref.~\cite{Mishra:PRE2010}. 

We perform direct numerical simulations (DNS) for Prandtl numbers $10^2, 10^3$, and $\infty$ and Rayleigh numbers in the range $10^5$ to $5 \times 10^8$.  The parameters and grid resolutions of all our runs are listed in Table~\ref{table:details}. Our grid resolution is such that the Batchelor length scale is larger than the mean grid spacing, thus  ensuring that our simulations are fully resolved.  Quantitatively,   $k_{\mathrm{max}} \eta_\theta \geqslant 1$ for all the runs, where $k_\mathrm{max}$ is the maximum wavenumber (inverse of the smallest length scale), and $\eta_\theta = (\kappa^3/\epsilon_u)^{1/4}$ is the Batchelor length.  

We also perform simulations for $\mathrm{Pr} = 10^2$ in a 2D box of aspect ratio one with no-slip boundary condition on all sides. We use the spectral element code NEK5000~\cite{Fischer:JCP1997}. The Rayleigh number is varied from $10^4$ to $5 \times 10^7$. We chose a box with $28 \times 28$ spectral elements and 7th-order polynomials within each element, therefore overall grid resolution is $196^2$. For the spectra study, however, we use 15th-order polynomials that yields $420^2$ effective grid points in the box.  

We compute the energy and entropy spectra and fluxes, Nusselt and P\'{e}clet numbers, temperature fluctuations and dissipation rates using the numerical data of the steady state. These quantities are averaged over 2000 eddy turnover times. 

\begin{table*}
\begin{ruledtabular}
\caption{Details of our free-slip numerical simulations: $N_x, N_y$ and $N_z$ are the number of grid points in $x$-, $y$-, and $z$-directions, respectively. The computed viscous dissipation rates $C_{\epsilon_u}^{\mathrm{comp.}}$ are in good agreement with the corresponding estimated values $C_{\epsilon_u}^{\mathrm{est.}}$ [= $\mathrm{(Nu-1)Ra/Pe}^2$]. Similarly, the computed thermal dissipation rates $C_{\epsilon_T,1}^{\mathrm{comp.}}$ and  $C_{\epsilon_T,2}^{\mathrm{comp.}}$  agree with the corresponding estimated values $C_{\epsilon_T,1}^{\mathrm{est.}} \, [= \mathrm{Nu}]$ and $C_{\epsilon_T,2}^{\mathrm{est.}} \, [= \mathrm{(Nu/Pe)}(\Delta/\theta_L)^2]$ reasonably well. For all the simulations $k_\mathrm{max} \eta_\theta \geqslant 1$, indicating that our simulations are well resolved.}
\begin{tabular}{ccccccccccc}
$\mathrm{Pr}$ & $\mathrm{Ra}$ & $N_x \times N_y \times N_z$ & Nu & Pe & $C_{\epsilon_u}^{\mathrm{comp.}}$ & $C_{\epsilon_u}^{\mathrm{est.}}$ & $C_{\epsilon_T,1}^{\mathrm{comp.}}$ & $C_{\epsilon_T,2}^{\mathrm{comp.}}$ & $C_{\epsilon_T,2}^{\mathrm{est.}}$ & $k_{\mathrm{max}}\eta_{\theta}$ \\
\hline
$10^2$ & $1 \times 10^5$ & 256 $\times$ 1 $\times$ 128 & 9.8 & $1.98 \times 10^2$  & 22.3 & 22.3 & 9.8 & 0.61 & 0.61 & 2.9 \\
$10^2$ & $5 \times 10^5$ & 256 $\times$ 1 $\times$ 128 & 14.5 & $4.98 \times 10^2$ & 28.5 & 27.3 & 14.5 & 0.37 & 0.35 & 1.8 \\
$10^2$ & $1 \times 10^6$ & 512 $\times$ 1 $\times$ 128 & 17.3 & $7.16 \times 10^2$ & 34.4 & 31.8 & 17.3 & 0.31 & 0.29 & 2.0 \\
$10^2$ & $5 \times 10^6$ & 512 $\times$ 1 $\times$ 256 & 27.4 & $1.84 \times 10^3$ & 42.5 & 38.9 & 27.4 & 0.19 & 0.18 & 1.7 \\
$10^2$ & $1 \times 10^7$ & 1024 $\times$ 1 $\times$ 256 & 34.7 & $3.13 \times 10^3$ & 36.5 & 34.5 & 34.7 & 0.14 & 0.13 & 1.9 \\
$10^2$ & $5 \times 10^7$ & 1024 $\times$ 1 $\times$ 512 & 61.6 & $1.03 \times 10^4$ & 28.7 & 28.6 & 61.6 & 0.072 & 0.072 & 1.5 \\
$10^2$ & $1 \times 10^8$ & 2048 $\times$ 1 $\times$ 512 & 79.8 & $1.70 \times 10^4$ & 27.1 & 27.1 & 79.8 & 0.056 & 0.056 & 1.7 \\

$10^3$ & $1 \times 10^5$ & 256 $\times$ 1 $\times$ 128 & 9.8 & $1.98 \times 10^2$  & 22.3 & 22.3 & 9.8 & 0.60 & 0.60 & 1.6 \\
$10^3$ & $5 \times 10^5$ & 512 $\times$ 1 $\times$ 128 & 16.0 & $5.36 \times 10^2$ & 26.1 & 26.1 & 16.0 & 0.36 & 0.36 & 1.4 \\
$10^3$ & $1 \times 10^6$ & 512 $\times$ 1 $\times$ 256 & 19.8 & $8.24 \times 10^2$ & 27.7 & 27.7 & 19.8 & 0.29 & 0.29 & 1.5 \\
$10^3$ & $5 \times 10^6$ & 1024 $\times$ 1 $\times$ 512 & 28.9 & $2.10 \times 10^3$ & 33.2 & 31.7 & 28.9 & 0.17 & 0.16 & 1.9 \\
$10^3$ & $1 \times 10^7$ & 1024 $\times$ 1 $\times$ 512 & 35.4 & $3.26 \times 10^3$ & 33.5 & 32.4 & 35.4 & 0.13 & 0.13 & 1.5 \\
$10^3$ & $5 \times 10^7$ & 2048 $\times$ 1 $\times$ 1024 & 57.7 & $8.79 \times 10^3$ & 38.0 & 36.7 & 57.3 & 0.080 & 0.078 & 1.7 \\

$\infty$ & $1 \times 10^5$ & 128 $\times$ 1 $\times$ 64 & 9.8 & $1.98 \times 10^2$ & 22.3 & 22.3 & 9.8 & 0.60 & 0.60 & 4.5 \\
$\infty$ & $5 \times 10^5$ & 128 $\times$ 1 $\times$ 64 & 16.0 & $5.37 \times 10^2$ & 26.1 & 26.1 & 16.1 & 0.36 & 0.36 & 2.7 \\
$\infty$ & $1 \times 10^6$ & 256 $\times$ 1 $\times$ 128 & 19.8 & $2.25 \times 10^2$ & 27.6 & 27.6 & 19.8 & 0.29 & 0.29 & 4.3 \\
$\infty$ & $5 \times 10^6$ & 512 $\times$ 1 $\times$ 128 & 32.6 & $2.27 \times 10^3$ & 30.8 & 30.8 & 32.6 & 0.17 & 0.17 & 3.6 \\
$\infty$ & $1 \times 10^7$ & 512 $\times$ 1 $\times$ 256 & 40.5 & $3.52 \times 10^3$ & 31.9 & 31.9 & 40.5 & 0.14 & 0.14 & 4.0 \\
$\infty$ & $5 \times 10^7$ & 1024 $\times$ 1 $\times$ 256 & 60.0 & $9.51 \times 10^3$ & 33.5 & 32.6 & 60.0 & 0.077 & 0.075 & 3.5 \\
$\infty$ & $1 \times 10^8$ & 1024 $\times$ 1 $\times$ 512 & 74.3 & $1.49 \times 10^4$ & 33.9 & 32.9 & 74.7 & 0.061 & 0.059 & 3.9 \\
$\infty$ & $5 \times 10^8$ & 2048 $\times$ 1 $\times$ 512 & 124 & $4.27 \times 10^4$ & 34.8 & 33.7 & 124 & 0.036 & 0.034 & 3.2 \\

$10^2$ & $1.0 \times 10^5$ &	 $256^3$ & 9.8 & $1.98 \times 10^2$ & 22.3 & 22.3 & 9.8 & 0.60 & 0.60 & 1.9 \\
$10^2$ & $6.5 \times 10^5$ & $256^3$ & 17.3 & $6.15 \times 10^2$ & 28.6 & 28.3 & 17.5 & 0.36 & 0.34 & 1.0 \\
$10^2$ & $2.0 \times 10^6$ & $512^3$ & 24.1 & $1.20 \times 10^3$ & 32.1 & 32.2 & 24.1 & 0.25 & 0.24 & 1.4 \\
$10^2$ & $5.0 \times 10^6$ & $512^3$ & 31.0 & $1.96 \times 10^3$ & 39.5 & 39.1 & 30.9 & 0.19 & 0.19 & 1.1 \\
$10^2$ & $1.0 \times 10^7$ & $1024^3$ & 38.1 & $2.92 \times 10^3$ & 43.7 & 43.4 & 38.2 & 0.16 & 0.16 & 1.7 \\
		
$10^3$ & $6.5 \times 10^4$ & $256^3$ & 8.6 & $1.53 \times 10^2$ & 21.4 & 21.4 & 8.6 & 0.69 & 0.68 & 1.3 \\
$10^3$ & $1.0 \times 10^5$ & $256^3$ & 9.8 & $1.98 \times 10^2$ & 22.3 & 22.3 & 9.8 & 0.60 & 0.60 & 1.1 \\
$10^3$ & $3.2 \times 10^5$ & $512^3$ & 14.1 & $3.98 \times 10^2$ & 27.2 & 27.1 & 14.1 & 0.42 & 0.43 & 1.5\\
$10^3$ & $2.0 \times 10^6$ & $1024^3$ & 24.3 & $1.10 \times 10^3$ & 38.7 & 38.3 & 24.3 & 0.26 & 0.26 & 1.6\\
$10^3$ & $6.0 \times 10^6$ & $1024^3$ & 34.2 & $2.13 \times 10^3$ & 43.4 & 43.7 & 34.2 & 0.19 & 0.19 & 1.1\\

$\infty$ & $7.0 \times 10^4$ & $128^3$ & 8.8 & $1.59 \times 10^2$ & 21.4 & 21.6 & 8.8 & 0.67 & 0.68 & 1.7 \\
$\infty$ & $3.2 \times 10^5$ & $128^3$ & 14.1 & $4.14 \times 10^2$ & 25.1 & 25.1 & 14.1 & 0.41 & 0.42 & 2.0 \\
$\infty$ & $6.5 \times 10^5$ & $128^3$ & 17.4 & $6.36 \times 10^2$ & 26.7 & 26.7 & 17.4 & 0.33 & 0.34 & 1.6 \\
$\infty$ & $3.9 \times 10^6$ & $256^3$ & 30.3 & $1.95 \times 10^3$ & 30.3 & 30.4 & 30.3 & 0.19 & 0.19 & 1.8 \\
$\infty$ & $6.5 \times 10^6$ & $256^3$ & 36.1 & $2.70 \times 10^3$ & 33.5 & 31.8 & 36.0 & 0.16 & 0.16 & 1.5 \\
$\infty$ & $9.8 \times 10^6$ & $256^3$ & 41.2 & $3.34 \times 10^3$ & 35.8 & 35.6 &  41.1 & 0.15 & 0.15 & 1.3 \\
$\infty$ & $1.0 \times 10^8$ & $512^3$ & 87.5 & $1.38 \times 10^4$ & 45.6 & 45.3 & 87.2 & 0.07 & 0.07 & 1.3 \\
\end{tabular}
\label{table:details}
\end{ruledtabular} 
\end{table*}

\section{Low wavenumber Fourier modes of 2D and 3D flows} \label{sec:similarity}
\begin{figure}
\includegraphics[scale=0.35]{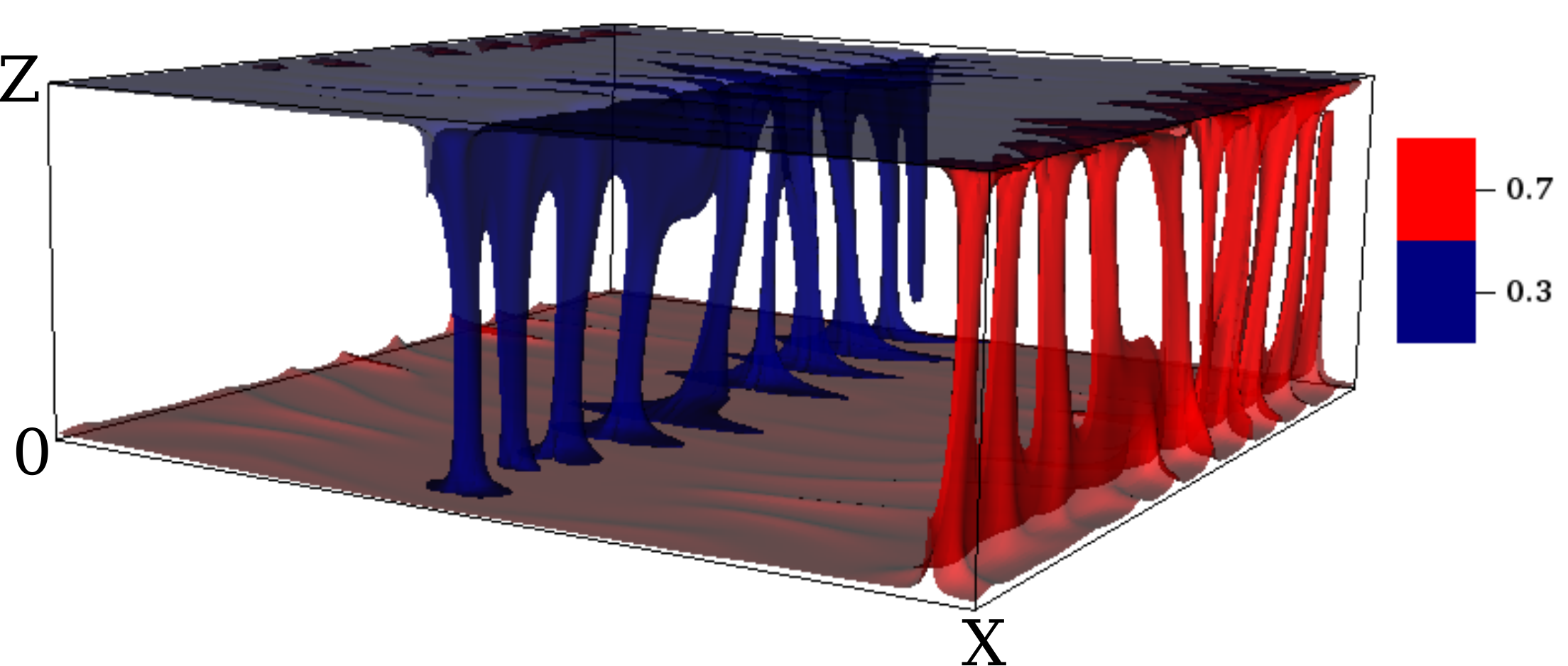}
\caption{Temperature isosurfaces for $\mathrm{Pr} = \infty$ and $\mathrm{Ra} = 6.6 \times 10^6$ exhibiting sharp plumes and quasi-2D nature of 3D RBC. The red (blue) structures represent hot (cold) fluid going up (down). [Figure adapted from Pandey \textit{et al.}~\cite{Pandey:PRE2014}]}
\label{fig:profile_3D}
\end{figure}

\begin{figure}
\includegraphics[scale=0.32]{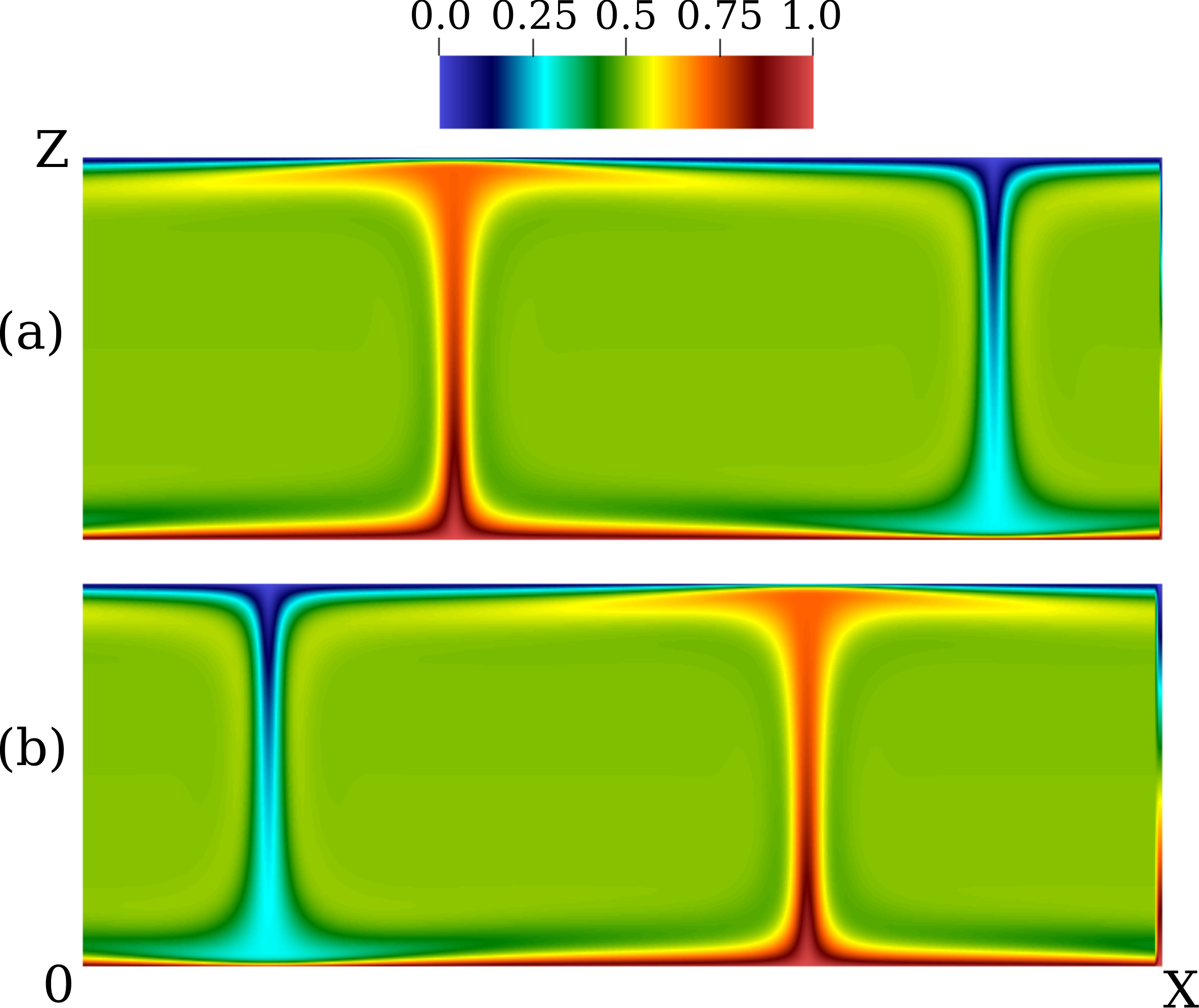}
\caption{Density plots of the temperature field in a 2D box for $\mathrm{Ra} = 10^6$ and  (a) $\mathrm{Pr} = 10^3$;  (b) $\mathrm{Pr} = \infty$. The figures illustrate hot (red) and cold (blue) plumes.}
\label{fig:profile}
\end{figure}

\begin{figure}
\includegraphics[scale=0.32]{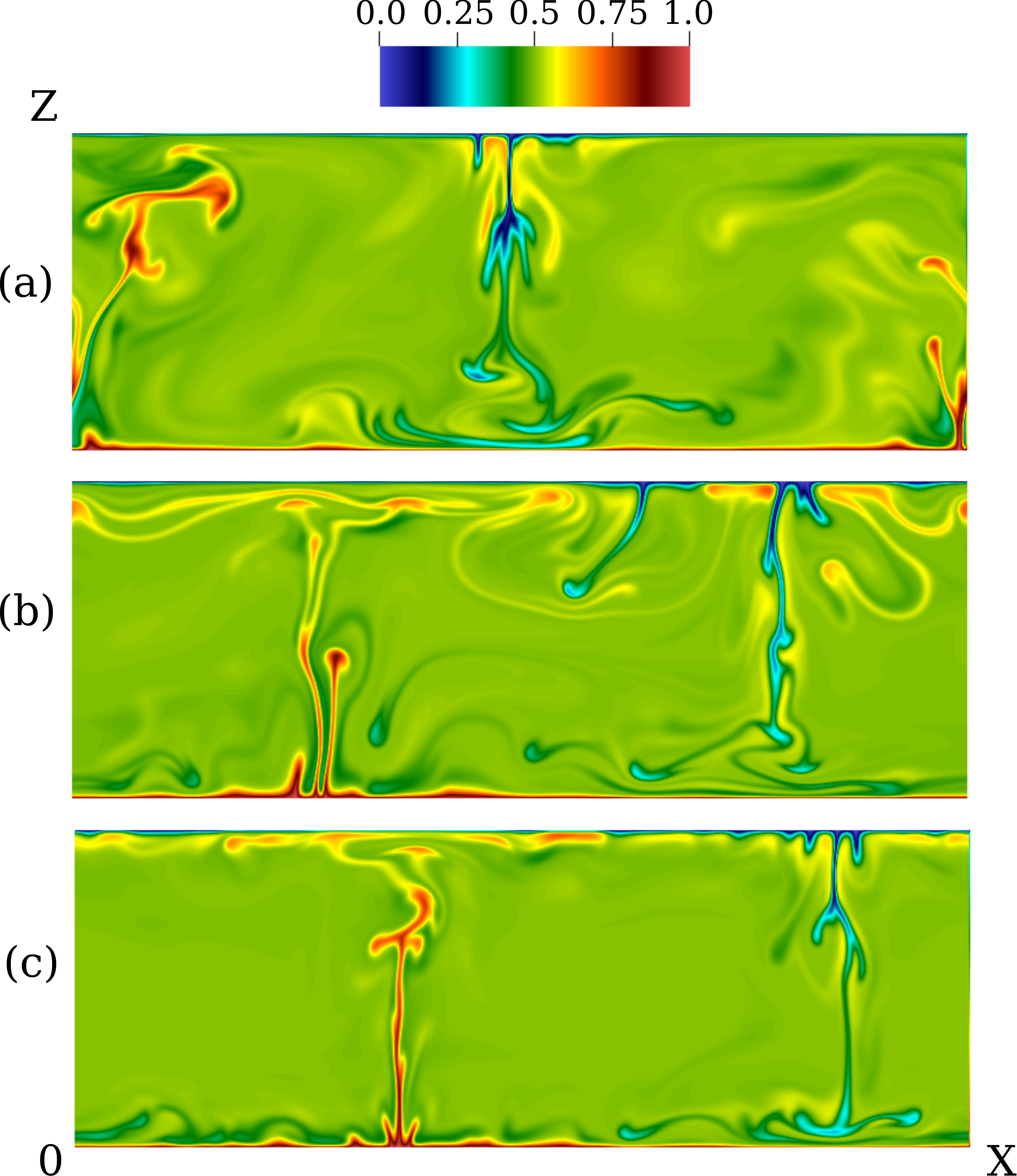}
\caption{Density plots of the temperature field for $\mathrm{Ra} = 5 \times 10^7$ and (a) $\mathrm{Pr} = 10^2$; (b) $\mathrm{Pr} = 10^3$; (c) $\mathrm{Pr} = \infty$.  The structures get sharper with increasing Prandtl numbers~\cite{Poel:JFM2013}.}
\label{fig:profile_ra5e7}
\end{figure}

Schmalzl \textit{et al.}~\cite{Schmalzl:GAFD2002, Schmalzl:EPL2004} and van der Poel \textit{et al.}~\cite{Poel:JFM2013} showed that the flow of  3D RBC resembles quite closely the 2D flow for large Prandtl numbers. The temperature isosurfaces for $\mathrm{Pr} = \infty$ shown in Fig.~\ref{fig:profile_3D} illustrates an array of parallel rolls, thus suggesting a quasi two-dimensional structures for the flow. For 2D RBC, the temperature field exhibited in Fig.~\ref{fig:profile} for $\mathrm{Pr} = 10^3, \infty$, and $\mathrm{Ra} =10^6$ resembles quite closely the rolls of 3D RBC. At larger Rayleigh numbers, the plumes become somewhat turbulent, as shown in  Fig.~\ref{fig:profile_ra5e7} for  $\mathrm{Ra} =5 \times 10^7$ and $\mathrm{Pr} = 100, 1000, \infty$.  Note that the plumes become sharper with increasing Prandtl number.   This similarity is because the most dominant $\theta$ modes are common among 2D and 3D RBC.  This is the reason why 3D RBC for large- and infinite Prandtl numbers is quasi two-dimensional.  

For comparison between the 2D and 3D RBC, we perform 2D and 3D simulations for $\mathrm{Pr} = \infty$ and $\mathrm{Ra} = 10^7$. In Table~\ref{table:energy}, we list the ten most dominant temperature modes along with their entropy and kinetic energy.  According to Table~\ref{table:energy},  the entropy and the kinetic  energy of the top ten modes,  $(k_x, k_z)$ in 2D and $(k_x, 0, k_z)$  in 3D, are very close.   This is the reason why the flow structures of the 3D RBC is quasi two-dimensional.  Also, the first six most dominant $\theta$ modes are $(0,0,2n) \approx -1/(2 n\pi)$, where $n=1..6$, as shown by Mishra and Verma~\cite{Mishra:PRE2010}; for these modes ${\mathbf u}({\mathbf k}) = 0$ [see Eq.~(\ref{eq:Eu_Eth})].  

\begin{table*}
\begin{ruledtabular}
\caption{Comparison of the ten most dominant entropy Fourier modes in 2D and 3D RBC for $\mathrm{Pr} = \infty$ and $\mathrm{Ra} = 10^7$. }
\begin{tabular}{cccccc}
Mode (3D) & $E_\theta^{\mathrm{mode}} / E^{3D}_\theta$  & $E_u^{\mathrm{mode}} / E^{3D}_u$ & Mode (2D) & $E_\theta^{\mathrm{mode}} / E^{2D}_\theta$ & $E_u^{\mathrm{mode}} / E^{2D}_u$ \\
($k_x, k_y, k_z$) & ($\%$) & ($\%$) & ($k_x, k_z$) & ($\%$) & ($\%$) \\
\hline
(0,0,2) & 30.4	& 0 & (0,2) & 30.6 & 0 \\
(0,0,4) & 7.81 & 0 & (0,4) & 7.79 & 0 \\
(0,0,6) & 3.56 & 0 & (0,6) & 3.54 & 0 \\
(0,0,8) & 2.03 & 0 & (0,8) & 2.03 & 0 \\
(0,0,10) & 1.30 & 0 & (0,10) & 1.31 & 0 \\
(0,0,12) & 0.87 & 0 & (0,12) & 0.90 & 0 \\
(1,0,1) & 0.018 & 19.6 & (1,1) & 0.020 & 20.4 \\
(3,0,1) & 0.011 & 2.05 & (3,1) & 0.018 & 3.39 \\
(1,0,3) & 0.011 & 0.046 & (1,3) & 0.011 & 0.046 \\
(3,0,3) & 0.003 & 0.039 & (3,3) & 0.007 & 0.095 \\
\end{tabular}
\label{table:energy}
\end{ruledtabular}
\end{table*}

Apart from $\hat{\theta}(0,0,2n)$ modes, the next four most dominant 2D modes are $(1,1), (3,1), (1,3)$, and $(3,3)$.  Clearly, $(1,1)$ is the most dominant mode with a finite kinetic energy, and it corresponds to a pair of rolls shown in Figs.~\ref{fig:profile_3D}$-$\ref{fig:profile_ra5e7}.  The mode $(1,1)$ is a part of the most dominant triad interaction $\{ (1,1), (-1,1), (0,2)\}$~\cite{Mishra:PRE2010}.  The other modes $ (3,1), (1,3)$ arise due to nonlinear interaction with the $(2,2)$ mode, which is relatively weak, but quite important~\cite{Chandra:PRL2013}.

We also compute the total energy of the three components of the velocity field in 3D, and the two components in 2D.  We observe that in 3D, $E_x/E_u = 0.55, E_y/E_u = 0.02$, and $E_z/E_u = 0.43$, clearly demonstrating the quasi 2D nature of the flow.  Here, $E_x = \langle u_x^2 \rangle /2$, $E_y = \langle u_y^2 \rangle /2$, $E_z = \langle u_z^2 \rangle /2$,  $E_u =  E_x+E_y+E_z$, and $\langle . \rangle$ represents time averaged value in the steady state.  In 2D, the ratios are $E_x/E_u = 0.58$ and $E_z/E_u = 0.42$, which are quite close to the corresponding ratios for the 3D RBC.    

We also performed similar analysis for $\mathrm{Pr}=100$ and 1000 for 2D and 3D, whose behaviour is similar to that for $\mathrm{Pr}=\infty$ described above.

Schmalzl \textit{et al.}~\cite{Schmalzl:GAFD2002, Schmalzl:EPL2004} decomposed the 3D velocity field into toroidal and poloidal components, and showed that the toroidal component disappears in the limit of infinite Prandtl number, consistent with the analytical results of Vitanov~\cite{Vitanov:PLA1998}. Schmalzl \textit{et al.}~\cite{Schmalzl:GAFD2002} argue that the vertical component of the vorticity disappears in the $\mathrm{Pr}=\infty$ limit, leading to vanishing of the toroidal component of the velocity, hence the two-dimensionalization of the $\mathrm{Pr}=\infty$ RBC.  Our results are consistent with those of Schmalzl \textit{et al.}~\cite{Schmalzl:GAFD2002, Schmalzl:EPL2004} and Vitanov~\cite{Vitanov:PLA1998}.
 
In the next section, we  will discuss the  spectra and fluxes of energy and entropy for large- and infinite Prandtl numbers.

\section{Energy spectra and fluxes} \label{sec:spectra}
In this section, we compute the  spectra and fluxes of energy and entropy for 2D and 3D RBC for large- and infinite Prandtl numbers and compare them.  We show that these quantities are very close to each other for 2D and 3D RBC because the dominant Fourier modes for them are very close to each other.

The one-dimensional kinetic energy and entropy spectra are defined as
\begin{eqnarray}
E_u(k) & = & \sum_{k \leq |{\bf k'}| < k+1} \frac{|\hat{\mathbf u}(\mathbf k')|^2}{2}, \label{eq:sptr_u} \\
E_{\theta}(k) & = & \sum_{k \leq |{\bf k'}| < k+1} \frac{|\hat{\theta}(\mathbf{k'})|^2}{2} \label{eq:sptr_th}.
\end{eqnarray}
The flow is  anisotropic in 2D RBC, e.g., $E_x/E_z = 1.37$, but the degree of anisotropy is rather small.  Hence, the aforementioned one-dimensional spectra are good description of the flow properties.

The nonlinear interactions induce kinetic energy and entropy transfers from larger length scales to smaller length scales that results in  kinetic energy and entropy fluxes.  Note that for  $\mathrm{Pr}=\infty$, the nonlinear interaction among the velocity modes  is absent, hence the kinetic energy flux is zero for this case.  The kinetic energy and entropy fluxes coming out of a wavenumber sphere of radius $k_0$ are given by~\cite{Verma:PR2004, Mishra:PRE2010}
\begin{equation}
\Pi_u(k_0) = \sum_{k \geq k_0} \sum_{p < k_0} \delta_{\mathbf k, \mathbf p+\mathbf q} \Im([\mathbf k \cdot \hat{\mathbf u}(\mathbf q)][\hat{\mathbf u}^*(\mathbf k) \cdot \hat{\mathbf u}(\mathbf p)]), \label{eq:u_flux}
\end{equation}
\begin{equation}
\Pi_{\theta}(k_0) = \sum_{k \geq k_0} \sum_{p < k_0} \delta_{\mathbf k, \mathbf p+\mathbf q} \Im([\mathbf k \cdot \hat{\mathbf u}(\mathbf q)][\hat{\theta}^*(\mathbf k) \cdot \hat{\theta} (\mathbf p)]), \label{eq:th_flux}
\end{equation} 
where $\Im$ stands for the imaginary part of the argument, and ${\mathbf k,\mathbf p,\mathbf q}$ are the wavenumbers of a triad with ${\mathbf k=\mathbf p+ \mathbf q}$. 

For 3D RBC with $\mathrm{Pr} = \infty$, Pandey \textit{et al.}~\cite{Pandey:PRE2014} derived the kinetic energy and entropy spectra as
\begin{eqnarray}
E_u(k) & = & (a_2^2  a_3)^{\frac{2}{3}} d \left(\frac{\kappa}{d}\right)^2 \mathrm{Ra}^{\frac{2}{3}(3-2\delta-\zeta)}(kd)^{-\frac{13}{3}}, \label{eq:Euk} \\
E_\theta(k) & = & (a_2^2  a_3)^{\frac{2}{3}} d \Delta^2 \mathrm{Ra}^{\frac{2}{3}(\delta-\zeta)} (kd)^{-\frac{1}{3}}, \label{eq:Ethk}
\end{eqnarray}
where $a_2$, $a_3$, $\zeta$ and $\delta$ are defined using $\theta_{\mathrm{rms}} = a_2 \Delta$, $\mathrm{Pe} = a_3 \mathrm{Ra}^{1 - \zeta}$, and  $\theta_{\mathrm{res}} \sim \mathrm{Ra}^{\delta}$. The $\theta_{\mathrm{res}}$ is the temperature fluctuation without $\hat{\theta}(0,0,2n)$ modes~\cite{Pandey:PRE2014}. They also argued that the kinetic energy flux $\Pi_u(k) \rightarrow 0$, but $\Pi_\theta(k) \approx \mathrm{const}$ in the inertial range for $\mathrm{Pr} = \infty$ RBC. They showed that the above formulae also describe the energy spectra for very large Prandtl numbers, e.g., for $\mathrm{Pr} > 100$.

The arguments of Pandey \textit{et al.}~\cite{Pandey:PRE2014} are independent of dimensionality, hence we expect the above expressions to hold in 2D as well for large- and infinite Prandtl numbers.  In fact, the similarities must be very close because of the identical dominant Fourier modes in 2D and 3D RBC (see Sec.~\ref{sec:similarity}).  To verify the above conjecture, we compute the energy and entropy spectra, as well as their fluxes.  

\begin{figure}
\includegraphics[scale=0.23]{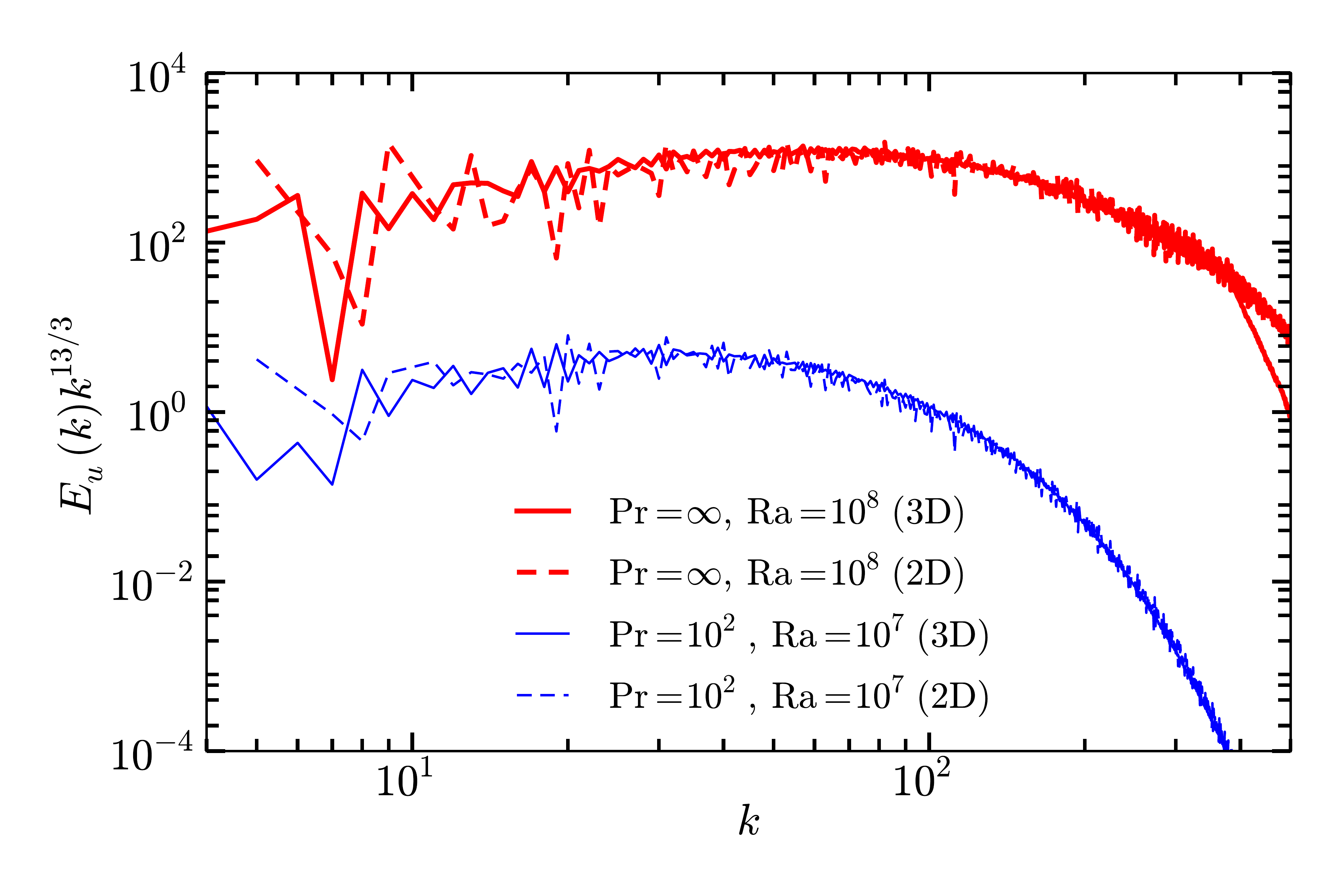}
\caption{The compensated kinetic energy spectrum $E_u(k)k^{13/3}$ as a function of wavenumber. Curves for 2D and 3D collapse on each other and are nearly constant in the inertial range, hence $E_u(k) \sim k^{-13/3}$.}
\label{fig:E_u}
\end{figure}

\begin{figure}
\includegraphics[scale=0.23]{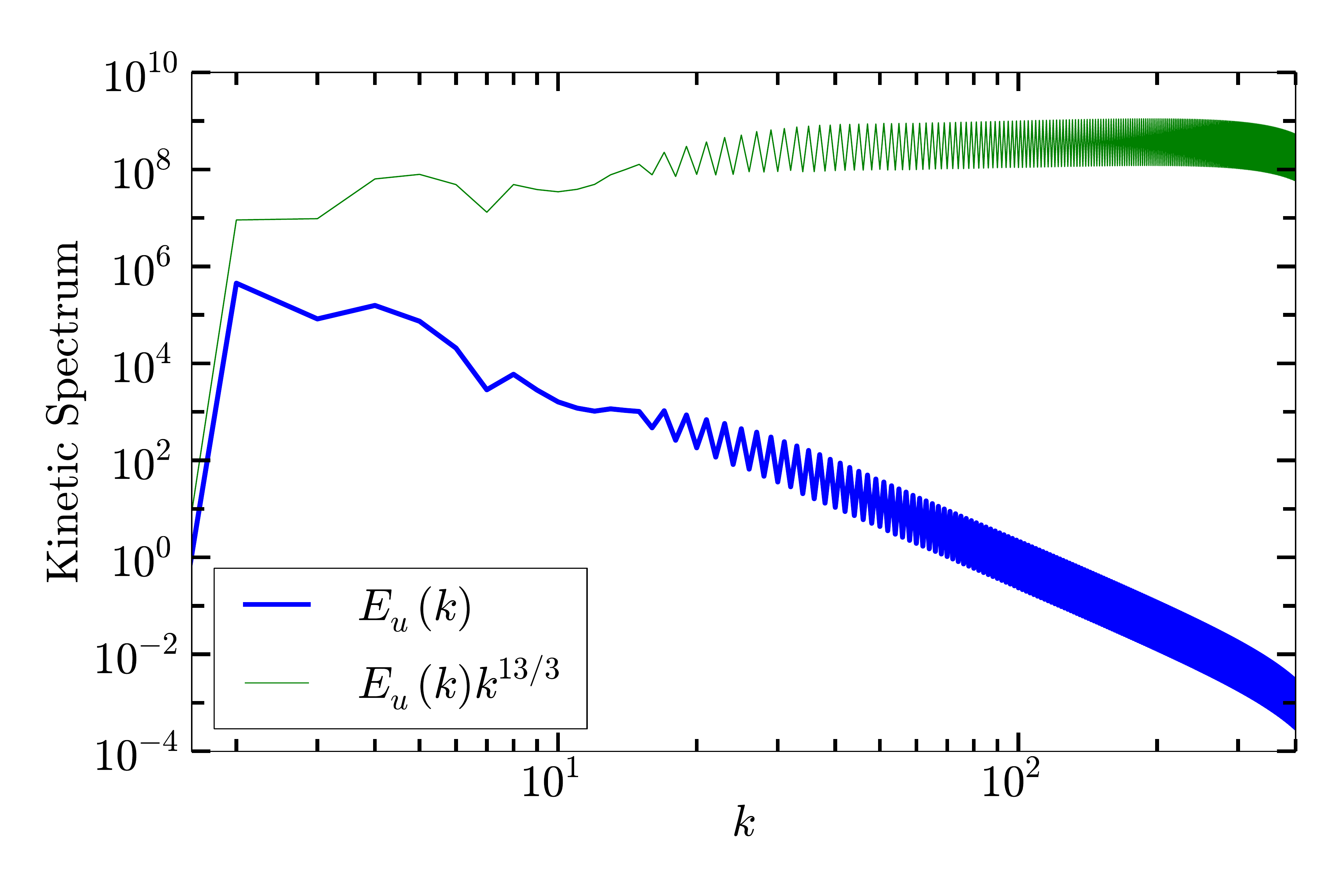}
\caption{The kinetic energy spectrum $E_u(k)$ for $\mathrm{Pr} = 10^2$ and $\mathrm{Ra} = 10^7$ with no-slip boundary condition in a 2D box. The normalized spectrum is nearly constant in the inertial range, hence $E_u(k) \sim k^{-13/3}$. [Figure adapted from Pandey \textit{et al.}~\cite{Pandey:PRE2014}]}
\label{fig:E_u_no}
\end{figure}

In Fig.~\ref{fig:E_u}, we plot the normalized kinetic spectrum $E_u(k)k^{13/3}$ for ($\mathrm{Pr} = 100, \mathrm{Ra} = 10^7$), and ($\mathrm{Pr} = \infty, \mathrm{Ra} = 10^8$) for both 2D and 3D RBC. The figure illustrates that the energy spectrum for 2D and 3D are quite close.  Hence, our conjecture that 2D and 3D RBC exhibit similar kinetic energy spectrum is verified.  Figure~\ref{fig:E_u_no} exhibits the kinetic spectrum for a RBC simulation in a unit box with no-slip boundary condition for  $\mathrm{Pr} = 100$ and $\mathrm{Ra} = 10^7$.  The figure demonstrates that $E_u(k) \sim k^{-13/3}$, similar to that of free-slip boundary condition.

\begin{figure}
\includegraphics[scale=0.23]{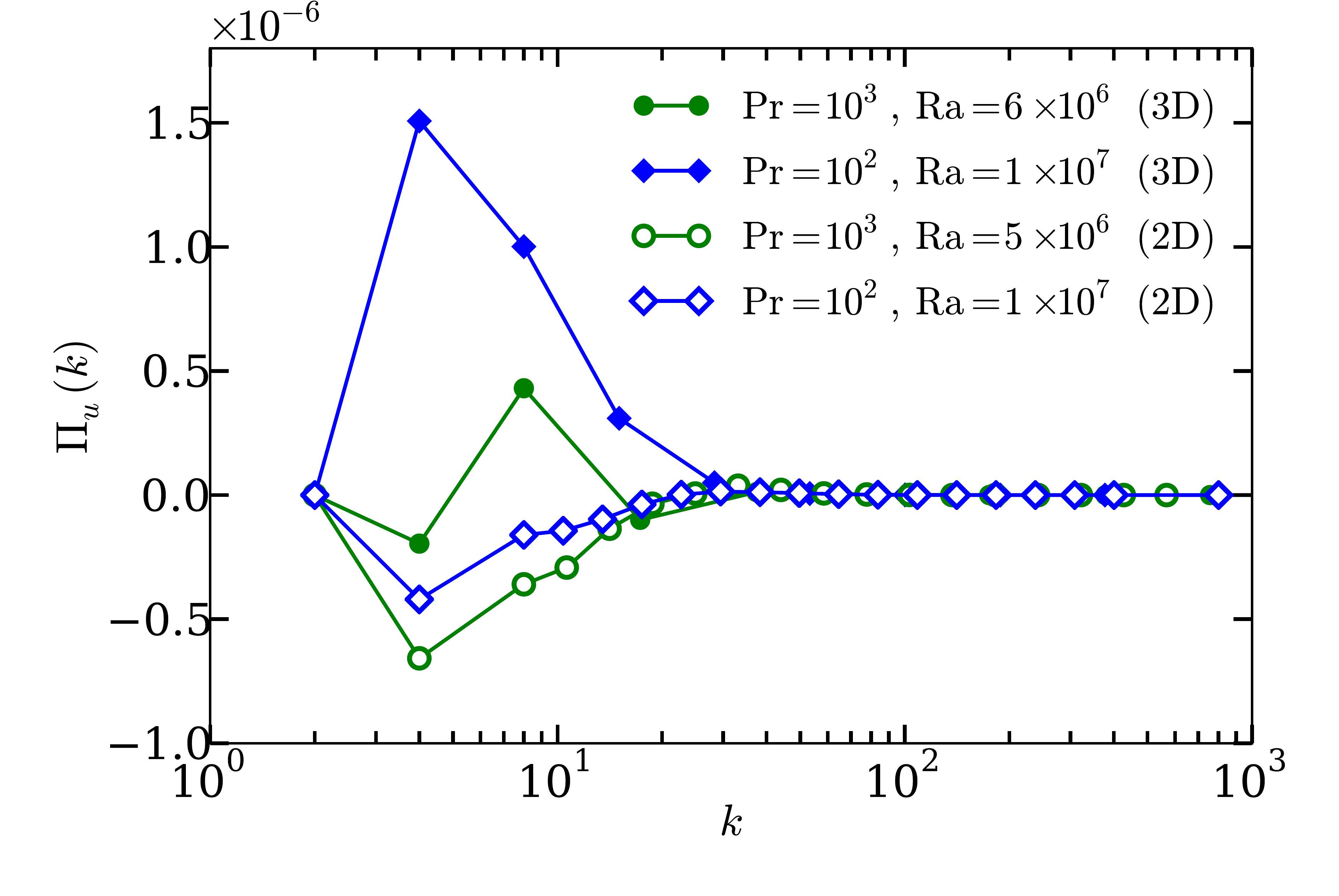}
\caption{Plot of the kinetic energy flux $\Pi_u(k)$ vs. $k$. The fluxes for $\mathrm{Pr} = 10^2$ have been multiplied by a factor of $10^{-2}$ to fit properly in this figure.  In 2D,  $\Pi_u(k)<0$, reminiscence of 2D fluid turbulence.}
\label{fig:Pi_u}
\end{figure}

The kinetic energy flux $\Pi_u$ for $\mathrm{Pr}=\infty$ is zero due to the absence of nonlinearity. However $\Pi_u$  is expected to be small ($\ll 1$ in normalized units of ours) for large $\mathrm{Pr}$.  In Fig.~\ref{fig:Pi_u}, we plot the kinetic energy flux $\Pi_u(k)$ for $\mathrm{Pr} = 10^2$ and  $10^3$ for 2D and 3D RBC.  As expected, the $\Pi_u$  are small for all the four cases. Interestingly, the kinetic energy flux for 2D RBC is negative at small wavenumbers, which is reminiscent of 2D fluid turbulence~\cite{Boffetta:ARFM2012, Toh:PRL1994}. The KE flux for 3D RBC is  positive almost everywhere. Thus, the KE fluxes for 2D and 3D RBC are somewhat different, but  they play insignificant role in the large- and infinite Prandtl number RBC.  Hence, we can claim that a common feature for the large $\mathrm{Pr}$ 2D and 3D RBC is that $\Pi_u \rightarrow 0$.  

\begin{figure}
\includegraphics[scale=0.23]{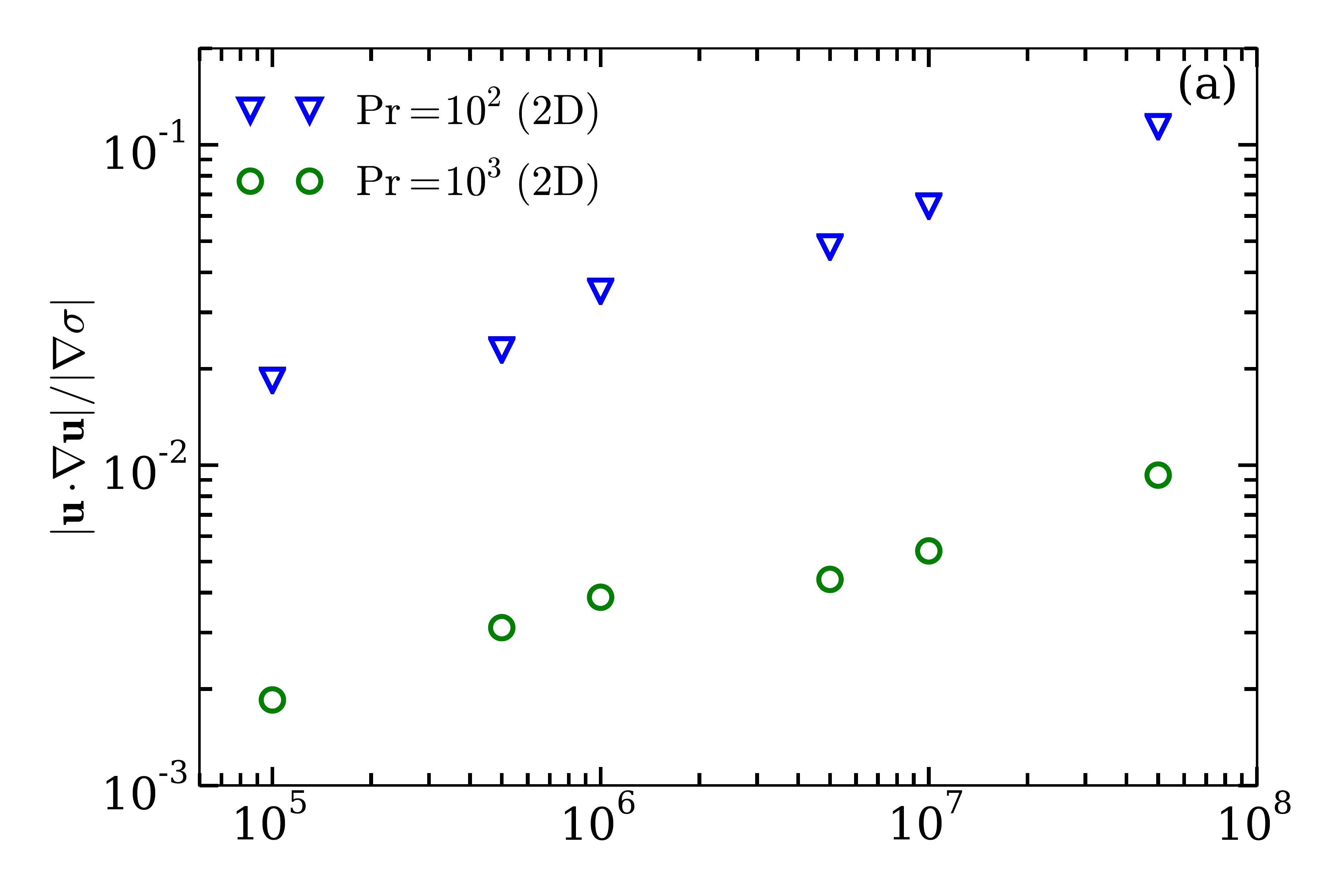}
\includegraphics[scale=0.23]{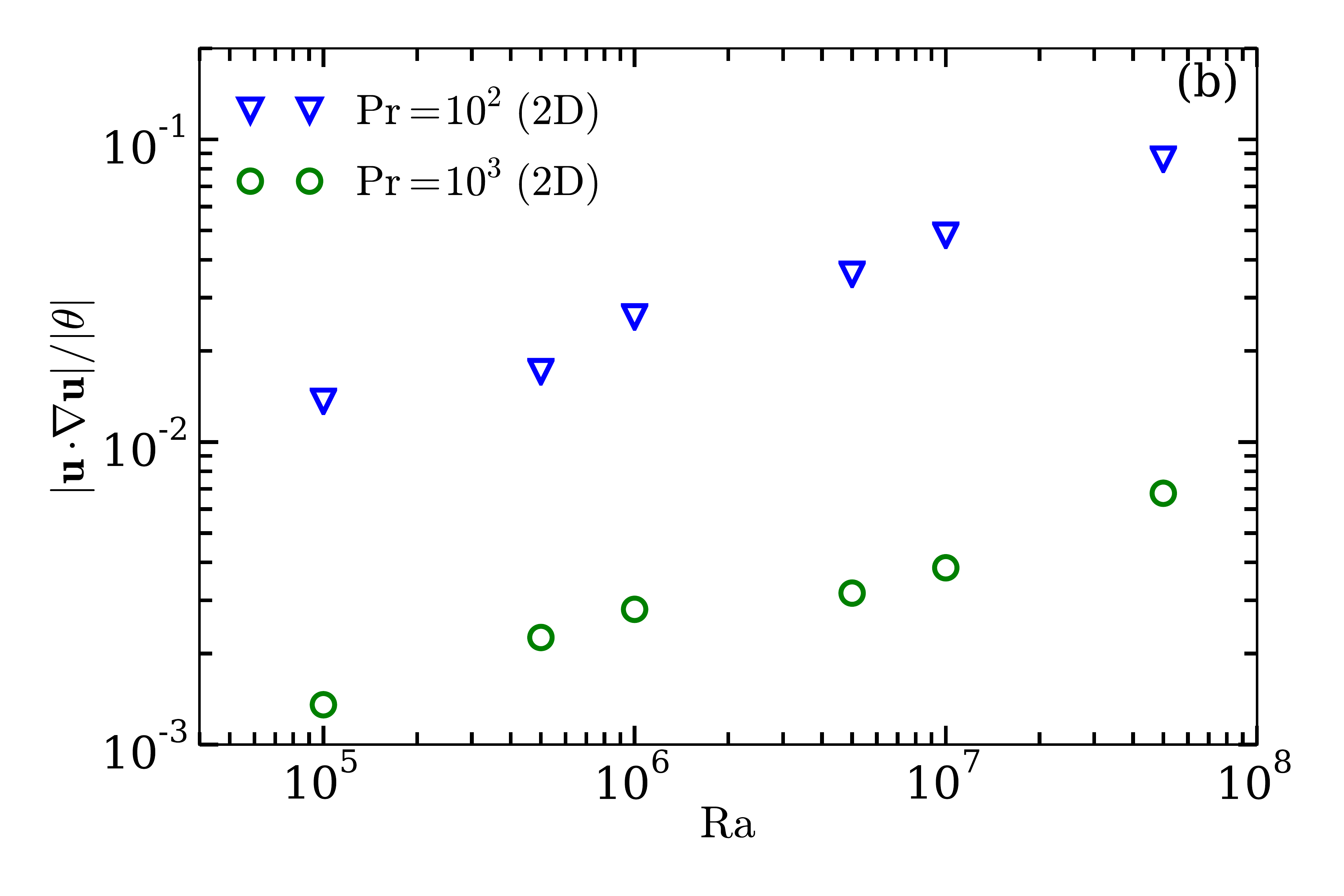}
\caption{Plots of the ratios between (a) nonlinear and pressure gradient terms; (b) nonlinear and buoyancy terms of Eq.~(\ref{eq:u_non}). The nonlinearity is weak compared to pressure gradient and buoyancy.} 
\label{fig:ratio}
\end{figure}

The smallness of kinetic energy flux for the large- and infinite $\mathrm{Pr}$ RBC is because the nonlinear term is much weaker than the pressure gradient and the buoyancy terms of Eq.~(\ref{eq:u_non}). In Fig.~\ref{fig:ratio}, we plot  $|({\bf u} \cdot \nabla){\bf u}|/|\nabla \sigma|$ and $|{\bf u} \cdot \nabla){\bf u}|/|\theta|$ as a function of $\mathrm{Ra}$.  The aforementioned  ratios lie between 0.001 to 0.1, and they become smaller as $\mathrm{Pr}$ increases.   These results show that the nonlinear term is weak for large- and infinite $\mathrm{Pr}$ RBC.  Note  that $|\nabla \sigma| \approx |\theta|$, consistent with Eq.~(\ref{eq:buoyancy_pressure}).

\begin{figure}
\includegraphics[scale=0.23]{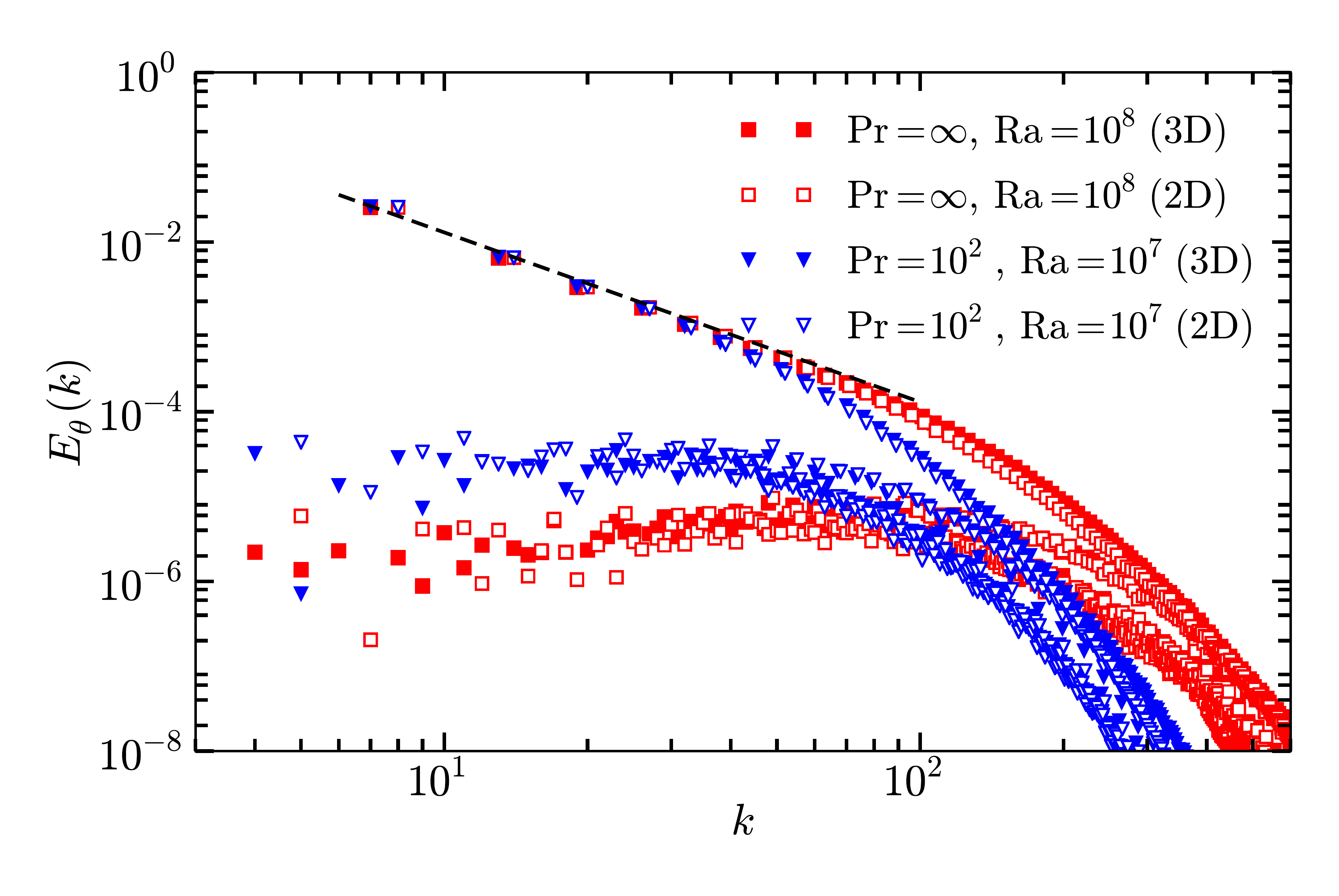}
\caption{Entropy spectrum $E_\theta(k)$ vs. $k$.  $E_\theta(k)$ exhibits a dual branch with a dominant upper branch with $E_\theta(k) \sim k^{-2}$.  The lower branch is almost flat in the inertial range. }
\label{fig:E_th}
\end{figure}

\begin{figure}
\includegraphics[scale=0.23]{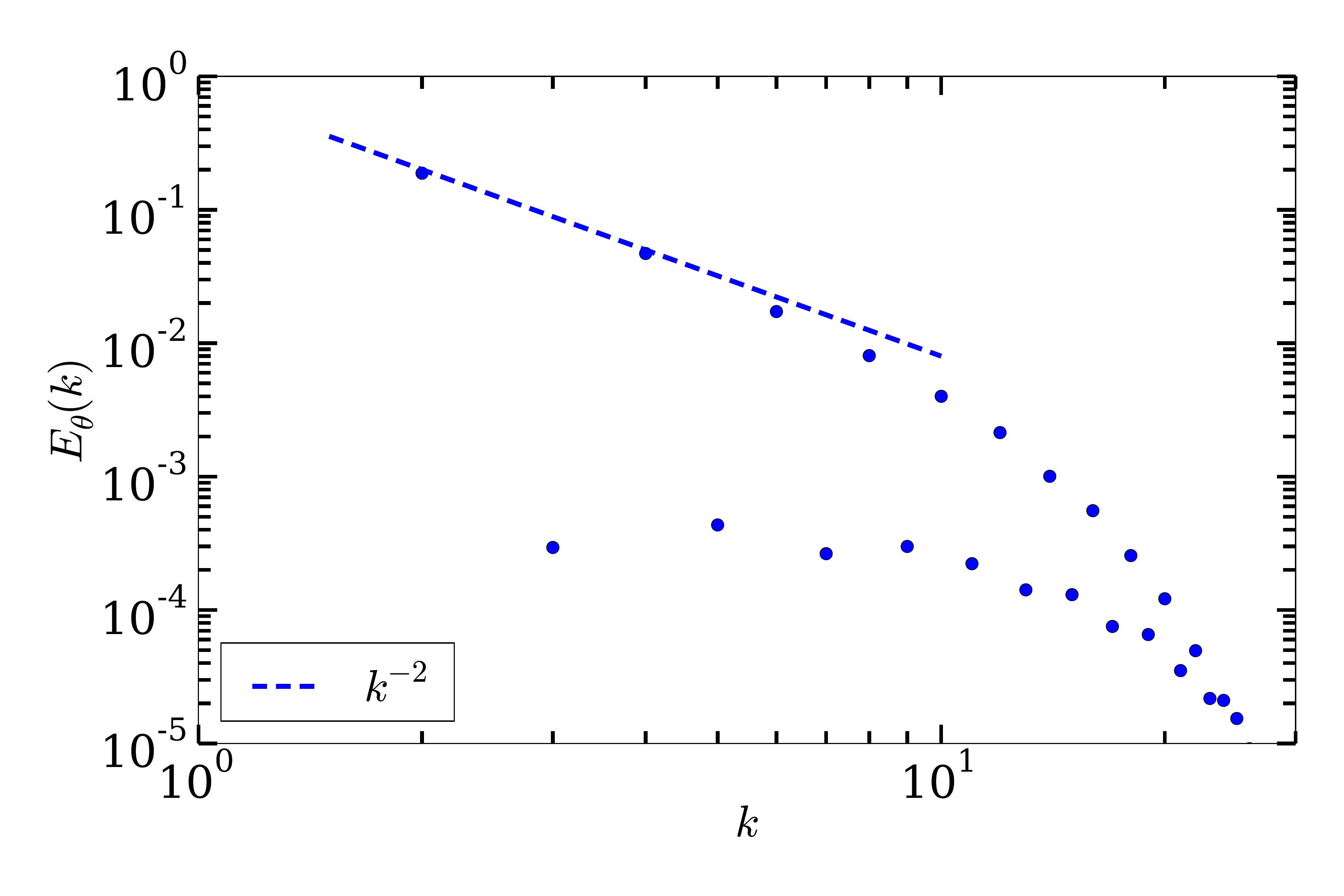}
\caption{Entropy spectrum $E_\theta(k)$ for $\mathrm{Pr} = 10^2$ and $\mathrm{Ra} = 10^7$ with no-slip boundary condition in a 2D box. It has a very similar behaviour as that for the free-slip boundary condition. [Figure adapted from Pandey \textit{et al.}~\cite{Pandey:PRE2014}]}
\label{fig:E_th_no}
\end{figure}

In Fig.~\ref{fig:E_th}, we plot the entropy spectrum for ($\mathrm{Pr}=100, \mathrm{Ra}=10^7$) and  ($\mathrm{Pr}=\infty, \mathrm{Ra}=10^8$) for 2D and 3D RBC.  Clearly, the  entropy spectrum for the 2D and 3D RBC also show very similar behaviour.  Note that the entropy spectrum exhibits a dual spectrum, with the top curve ($E(k) \sim k^{-2}$) representing the $\hat{\theta}(0,0,2n)$ modes, whose values are close to $-1/(2 n\pi)$ (see Sec.~\ref{sec:similarity} and Mishra and Verma~\cite{Mishra:PRE2010}).  The lower curve in the spectrum, corresponding to the modes other than $\hat{\theta}(0,0,2n)$, is somewhat flat.  We also observe similar entropy spectrum for no-slip boundary condition, which is shown in Fig.~\ref{fig:E_th_no} for $\mathrm{Pr} = 100$ and $\mathrm{Ra} = 10^7$.

\begin{figure}
\includegraphics[scale=0.23]{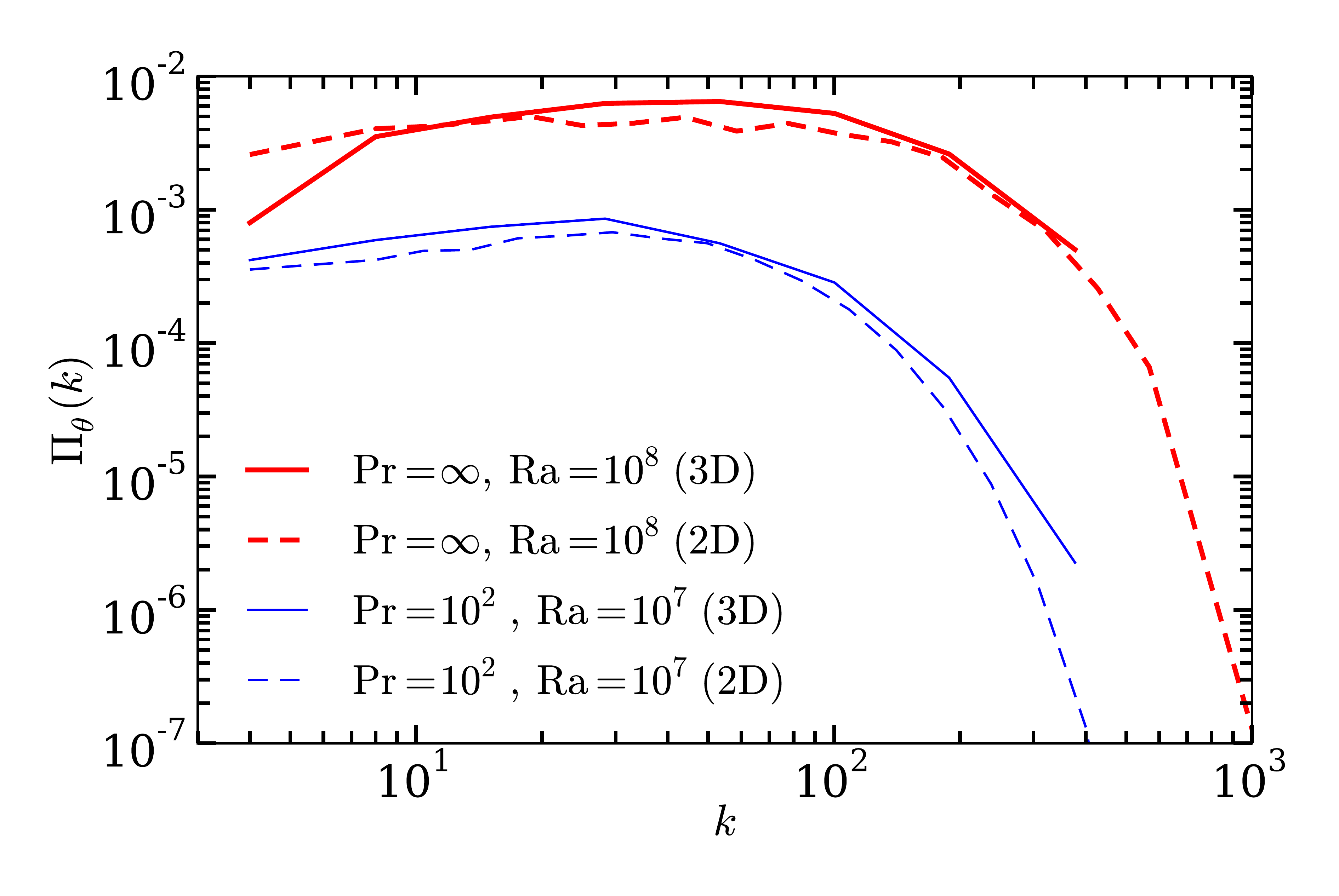}
\caption{Plot of the entropy flux $\Pi_\theta(k)$ vs.~$k$. The fluxes are nearly constant in the inertial range, and are similar for the 2D and 3D RBC.}
\label{fig:Pi_th}
\end{figure}

We  compute the entropy flux  defined in Eq.~(\ref{eq:th_flux})~\cite{Mishra:PRE2010} for ($\mathrm{Pr} = 100$, $\mathrm{Ra} = 10^7$), and ($\mathrm{Pr} = \infty$, $\mathrm{Ra} = 10^8$) for both 2D and 3D RBC. In Fig.~\ref{fig:Pi_th}, we plot the entropy flux $\Pi_\theta(k)$ for the above four cases.  Clearly, the behaviour of 2D and 3D RBC are very similar, with a constant entropy flux in the inertial range. 

In the next section, we will compute the large-scale quantities for 2D and 3D RBC with large- and infinite Prandt numbers.

\section{Scaling of large-scale quantities} \label{sec:results}

\subsection{Nusselt and P\'{e}clet numbers}
Schmalzl \textit{et al.}~\cite{Schmalzl:GAFD2002, Schmalzl:EPL2004} and van der Poel \textit{et al.}~\cite{Poel:JFM2013} showed that the Nusselt and P\'{e}clet numbers for 2D and 3D RBC exhibit similar scaling.  For validation of our data, we also compute the Nusselt  number $\mathrm{Nu}$ and  P\'{e}clet number   $\mathrm{Pe}$, as well as $\theta_\mathrm{rms}$ using our data sets.

\begin{figure}
\includegraphics[scale=0.23]{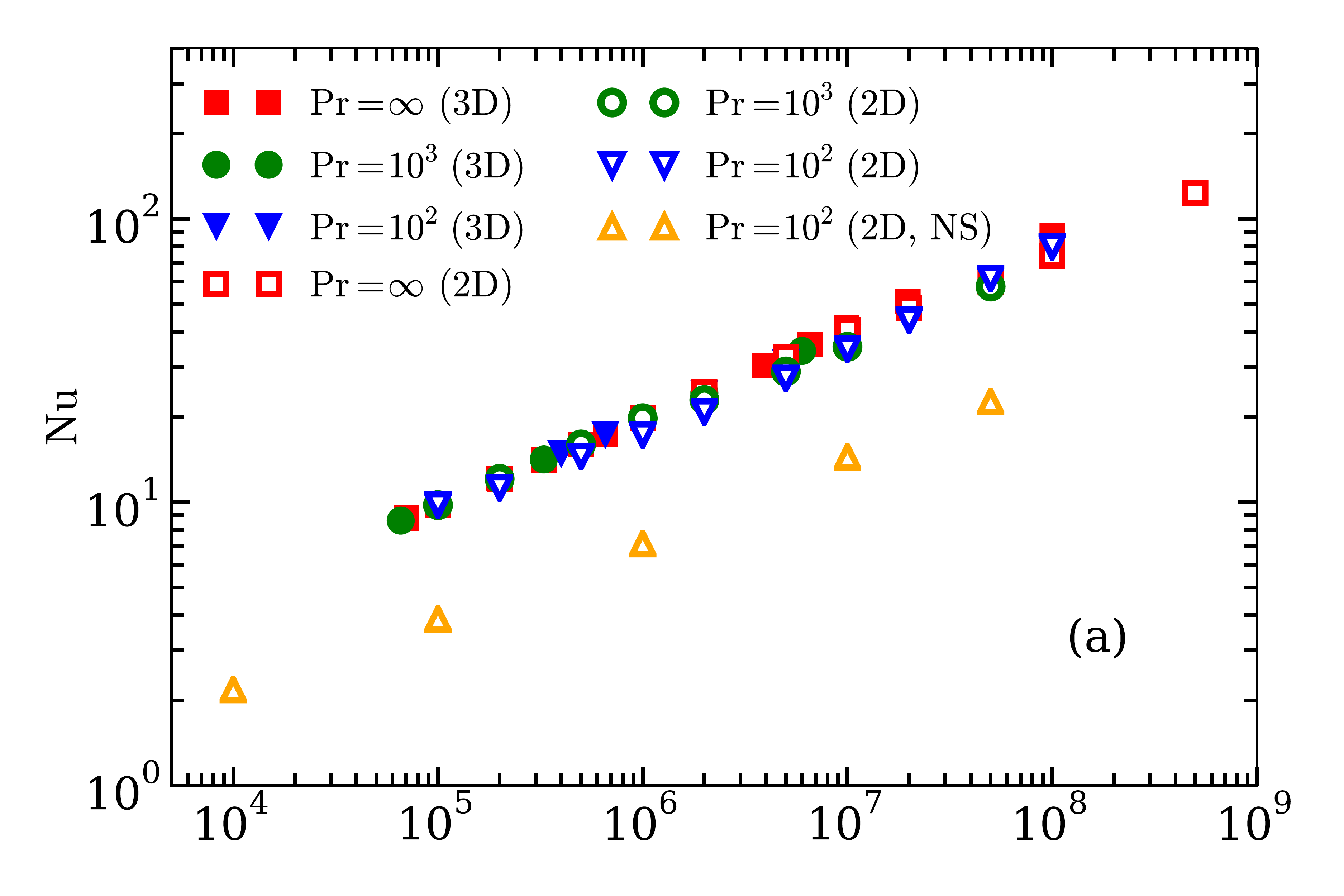}
\includegraphics[scale=0.23]{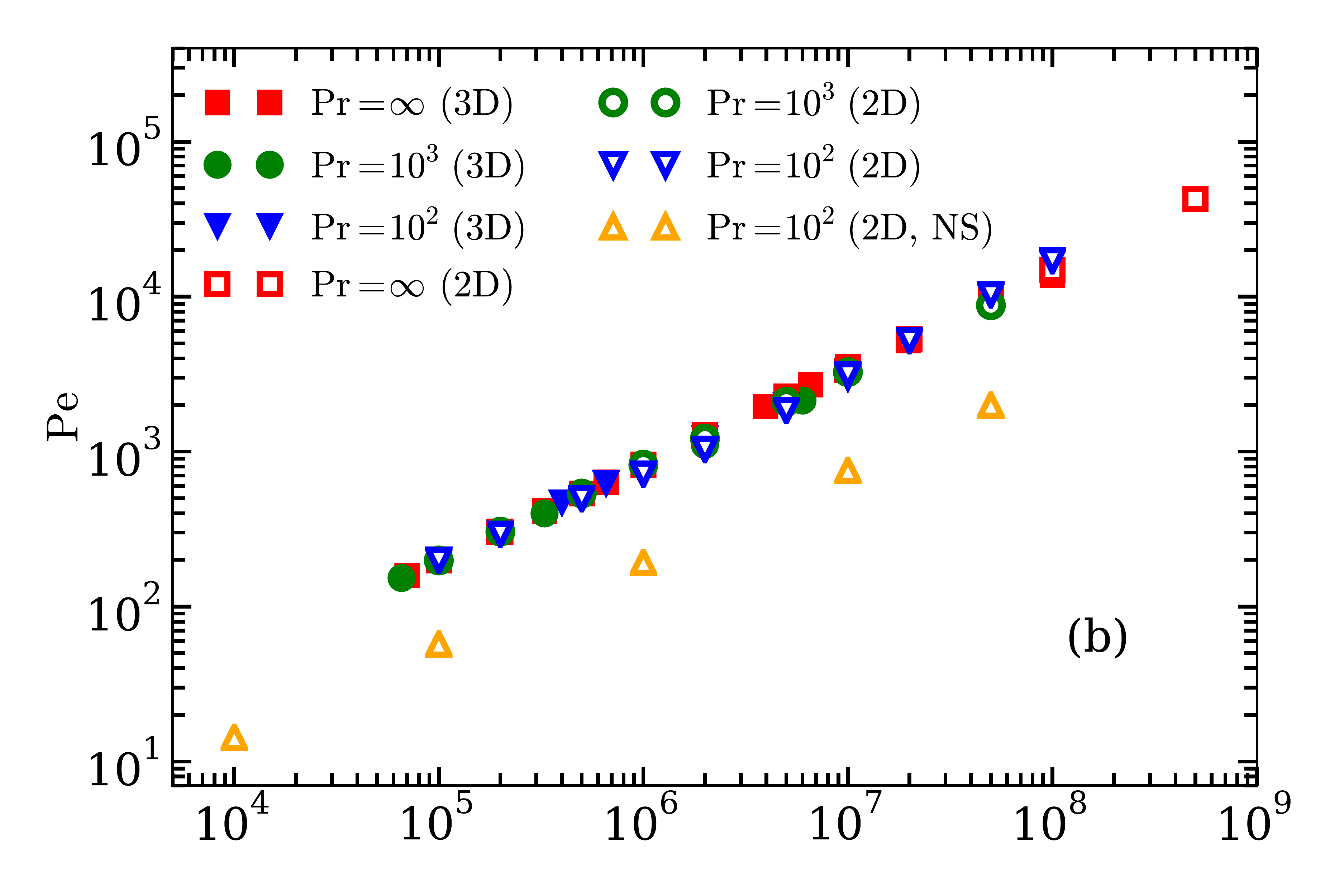}
\includegraphics[scale=0.23]{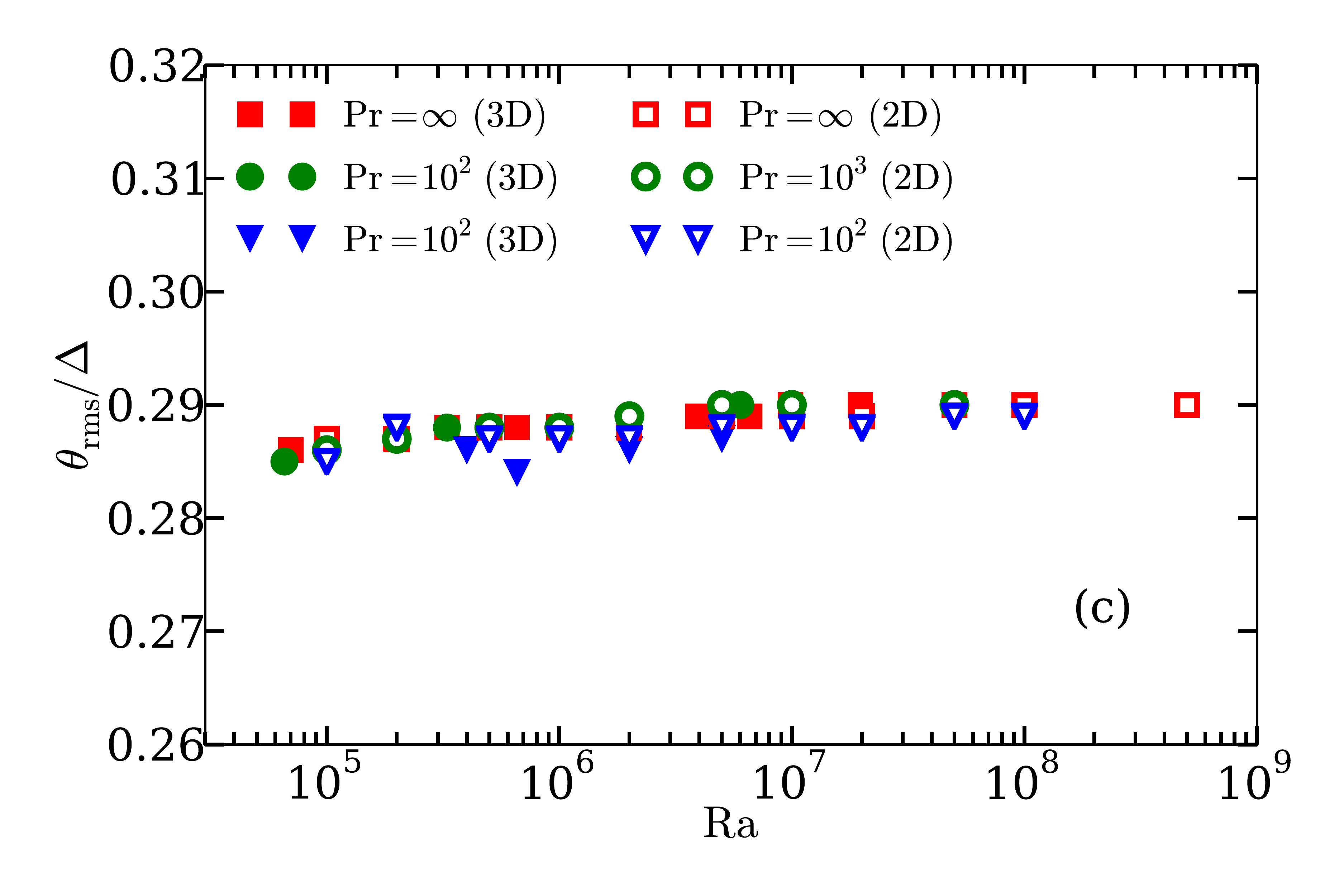}
\caption{Plots of (a) Nusselt number $\mathrm{Nu}$; (b) P\'{e}clet number $\mathrm{Pe}$; (c) normalized root mean square temperature fluctuations $\theta_{\mathrm{rms}}/\Delta$ as a function of Rayleigh number. The 2D and 3D RBC exhibit similar scaling for large-scale quantities, except for the no-slip data for $\mathrm{Pr} = 100$ (orange triangles), for which the prefactors are lower.}
\label{fig:nu_pe}
\end{figure}

In Fig.~\ref{fig:nu_pe}, we plot the Nusselt number, P\'{e}clet number and normalized root mean square temperature fluctuations for $\mathrm{Pr}=100, 1000, \infty$ and $\mathrm{Ra}$ ranging from $10^4$ to $5 \times 10^8$  for both 2D and 3D RBC. We also plot $\mathrm{Nu}$ and $\mathrm{Pe}$ for $\mathrm{Pr} = 100$ with no-slip boundary condition (shown by orange triangles). The figures show that the 2D and 3D RBC have similar Nusselt and P\'{e}clet number scaling,  in particular $\mathrm{Nu} \sim \mathrm{Ra}^{0.3}$ and $\mathrm{Pe}  \sim  \mathrm{Ra}^{0.6}$, with a weak variation of the exponents with $\mathrm{Pr}$ and $\mathrm{Ra}$.   However, the $\mathrm{Nu}$ and $\mathrm{Pe}$ prefactors for the no-slip data are lower than those for free-slip runs, which is due to lower frictional force for the free-slip boundary condition. These results are consistent with those of Schmalzl \textit{et al.}~\cite{Schmalzl:GAFD2002, Schmalzl:EPL2004}, van der Poel \textit{et al.}~\cite{Poel:JFM2013}, Silano \textit{et al.}~\cite{Silano:JFM2010}, and Pandey \textit{et al.}~\cite{Pandey:PRE2014}.  

We observe that $\theta_\mathrm{rms}/\Delta$ is a constant.   The details of scaling and error bars are discussed in Pandey \textit{et al.}~\cite{Pandey:PRE2014}.  The above similarities are primarily due to the quasi 2D nature of the 3D RBC.

\subsection{Dissipation rates}
In this subsection, we will discuss the scaling of normalized viscous and thermal dissipation rates for large Prandt numbers. Shraiman and Siggia~\cite{Shraiman:PRA1990}  derived the following exact relations between dissipation rates, $\mathrm{Pr}$, $\mathrm{Ra}$, and $\mathrm{Nu}$:
 \begin{eqnarray}
\epsilon_u & = & \nu \langle |\nabla \times \mathbf u|^2 \rangle = \frac{\nu^3}{d^4} \frac{(\mathrm{Nu}-1)\mathrm{Ra}}{\mathrm{Pr}^2}, \label{eq:eps_u} \\
\epsilon_T & = & \kappa \langle |\nabla T|^2 \rangle =  \kappa \frac{\Delta^2}{d^2} \mathrm{Nu}, \label{eq:eps_th}  
\end{eqnarray}
where $\epsilon_u$ and $\epsilon_T$ are the volume-averaged viscous and thermal dissipation rates, respectively.  For large- and infinite  Prandtl numbers,  which corresponds to the viscous dominated regime, an appropriate formula for the normalized viscous dissipation rate is~\cite{Pandey:PRE2014}
\begin{equation}
C_{\epsilon_u} = \frac{\epsilon_u}{\nu U_L^2/d^2} = \frac{(\mathrm{Nu}-1)\mathrm{Ra}}{\mathrm{Pe}^2}. \label{eq:C_eps_u2}
\end{equation}
The corresponding formulas for the normalized thermal dissipation rate are
\begin{eqnarray}
\label{eq:C_eps_t1} 
C_{\epsilon_T,1} & = & \frac{\epsilon_T}{\kappa \Delta^2/d^2} = \mathrm{Nu}, \label{eq:C_eps_t1} \\
C_{\epsilon_T,2} & = & \frac{\epsilon_T}{U_L \theta_L^2/d} = \frac{\mathrm{Nu}}{\mathrm{Pe}} \left( \frac{\Delta}{\theta_L} \right)^2. \label{eq:C_eps_t2}
\end{eqnarray}
See Pandey \textit{et al.}~\cite{Pandey:PRE2014} for a detailed discussion on the dissipation rates for large Prandtl number convection.

\begin{figure}
\includegraphics[scale=0.23]{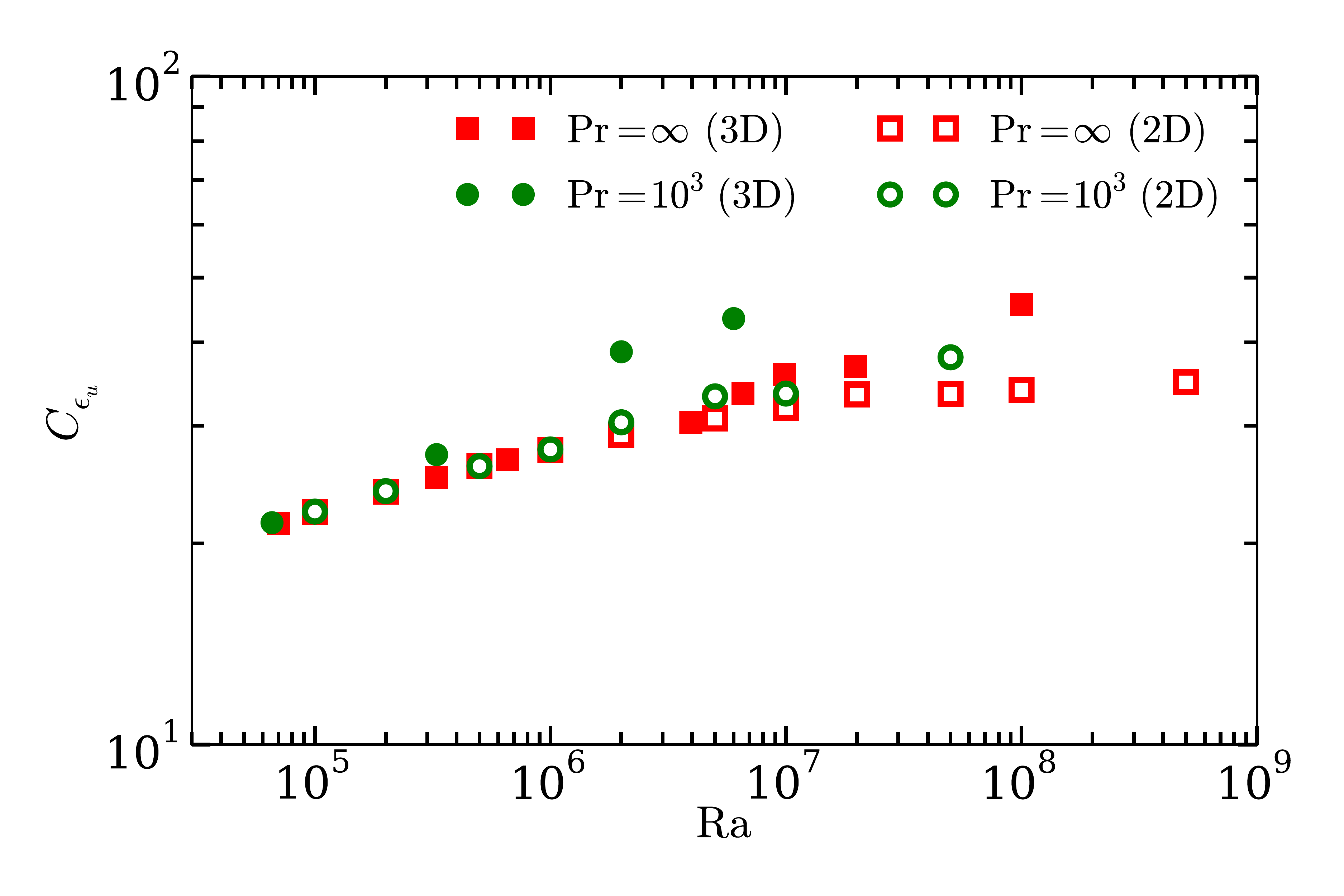}
\caption{Normalized viscous dissipation rate $C_{\epsilon_u}$ as a function of $\mathrm{Ra}$. The values of $C_{\epsilon_u}$ are lower in 2D compared to 3D RBC.}
\label{fig:c_eps_u}
\end{figure}

Using the scaling of Nu and Pe, we find that  for $\mathrm{Pr} = \infty$, $C_{\epsilon_u}$ is an approximate constant independent of $\mathrm{Ra}$~\cite{Pandey:PRE2014}. In Fig.~\ref{fig:c_eps_u}, we plot $C_{\epsilon_u}$ for $\mathrm{Pr} = 10^3$ and $\infty$, according to which  $C_{\epsilon_u}$ is nearly a constant with a significant scatter of data.   As evident from the figure, the normalized viscous dissipation rate for the 2D RBC is a bit lower than the corresponding data for the 3D RBC, which is due the inverse cascade of energy in 2D RBC that suppresses $\Pi_u$ (see Fig.~\ref{fig:Pi_u}).

In Table 1, we list the normalized thermal dissipation rate $C_{\epsilon_T,1}$ and the Nusselt number, and they are observed to be quite close to each other, consistent with Eq.~(\ref{eq:C_eps_t1}). In the table, we also list the computed dissipation rate $C_{\epsilon_T,2}^\mathrm{comp.} = \epsilon_T/ (U_L \theta_L^2/d)$ and the estimated dissipation rate $C_{\epsilon_T,2}^\mathrm{est.} = \mathrm{(Nu/Pe)}(\Delta/\theta_L)^2$,  where $U_L  = \sqrt{2 E_u}$ and $ \theta_L = \sqrt{2 E_\theta}$.  These quantities are close to  each other, consistent with Eq.~(\ref{eq:C_eps_t2})

\begin{figure}
\includegraphics[scale=0.23]{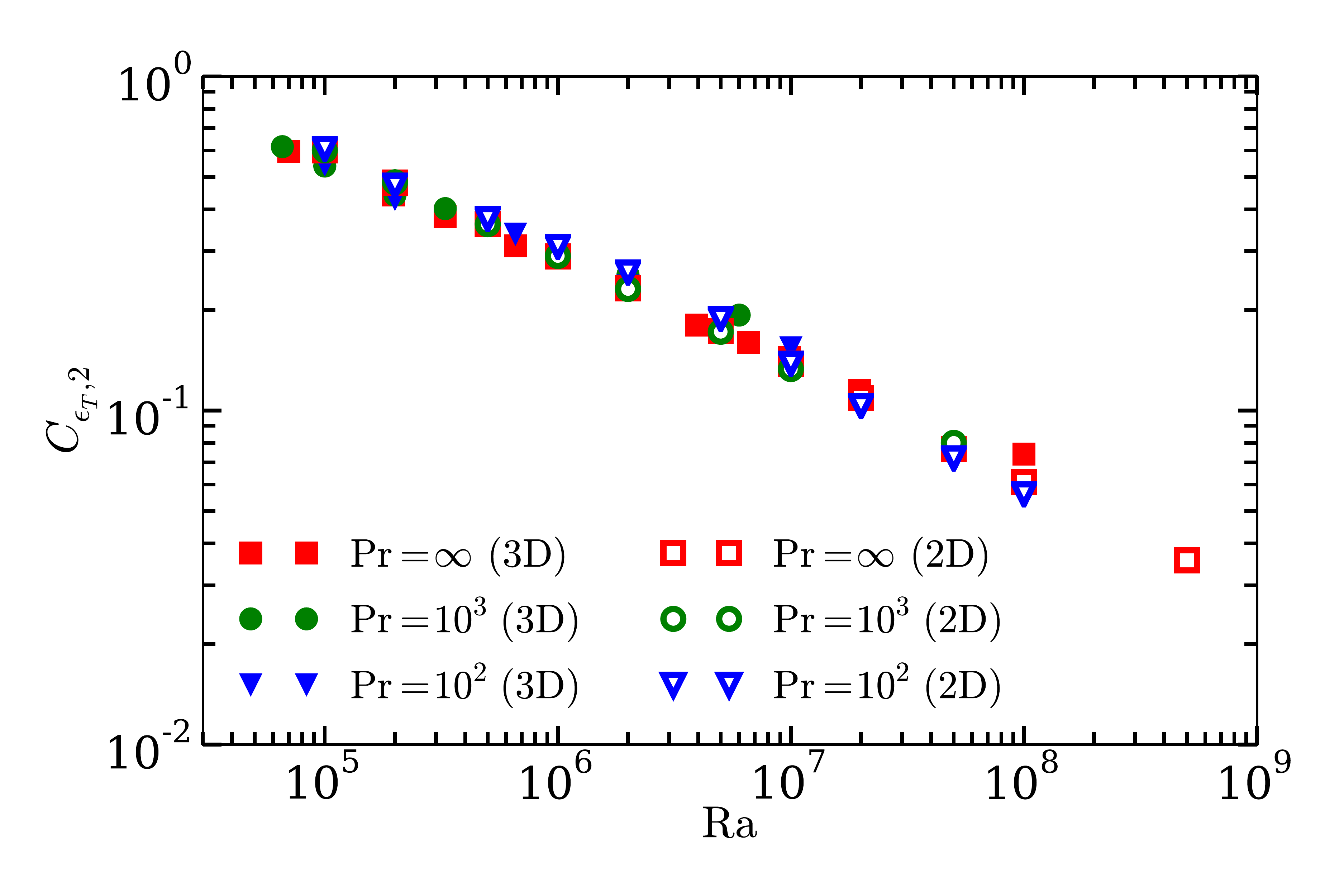}
\caption{Normalized thermal dissipation rate $C_{\epsilon_T,2}$ as a function of Rayleigh number. We observe $C_{\epsilon_T,2} \sim \mathrm{Ra}^{-0.32}$ in 2D, which is similar to the scaling for $\mathrm{Pr} = \infty$ in 3D.}
\label{fig:c_eps_t2}
\end{figure}

Figure~\ref{fig:c_eps_t2} exhibits $C_{\epsilon_T,2}$ as a function of Ra. The figure shows that  the scaling of $C_{\epsilon_T,2}$ in 2D is  similar to that for 3D RBC.  A detailed analysis indicates that  for 2D RBC, $C_{\epsilon_T,2} = (22 \pm 9)\mathrm{Ra}^{-0.31 \pm 0.03}$,  $(24 \pm 1.7)\mathrm{Ra}^{-0.32 \pm 0.01}$ and $(24 \pm 2.1)\mathrm{Ra}^{-0.32 \pm 0.01}$ for $\mathrm{Pr} = 10^2$, $10^3$, and $\infty$, respectively.   

These computations show that the behaviour of viscous and thermal dissipation rates for 2D and 3D RBC are quite similar.

\section{Discussions and conclusions} \label{sec:conclusion}
We performed numerical simulations of 2D and 3D RBC for $\mathrm{Pr}=100, 1000, \infty$, and $\mathrm{Ra}$ in the range of $10^5$ to $5\times 10^8$.  We showed that the dominant Fourier modes of the 2D and 3D flows are very close to each other, which is the reason for the similarities between the Nusselt and P\'{e}clet numbers in 2D and 3D RBC, as reported by Schmalzl \textit{et al.}~\cite{Schmalzl:GAFD2002, Schmalzl:EPL2004} and van der Poel \textit{et al.}~\cite{Poel:JFM2013}.  The flow in 3D RBC is quasi two-dimensional because of the strong suppression of the velocity in one of the horizontal directions.  These results are consistent with the results of Schmalzl \textit{et al.} ~\cite{Schmalzl:GAFD2002, Schmalzl:EPL2004} and  Vitanov~\cite{Vitanov:PLA1998}, according to which the  toroidal component of the velocity field in 3D RBC vanishes for $\mathrm{Pr}=\infty$.

We  compute the spectra and fluxes of the kinetic energy and entropy for the 2D RBC, show them to be very similar to those for 3D RBC.  In particular, we observe that the kinetic energy spectrum $E_u(k) \sim k^{-13/3}$, while the entropy spectrum exhibits a dual branch, with a dominant $k^{-2}$ branch corresponding to the $\hat{\theta}(0,0,2n)$ Fourier modes.   The other entropy branch is somewhat flat.  The similarities between the spectra and fluxes of 2D and 3D RBC are due to the quasi 2D nature of 3D RBC.    

We compute the global quantities such as the Nusslet and P\'{e}clet numbers, $\theta_\mathrm{rms}$, the kinetic energy and thermal dissipation rates.   All these quantities exhibit similar behaviour in 2D and 3D RBC, which is consistent with the results of Schmalzl \textit{et al.}~\cite{Schmalzl:GAFD2002, Schmalzl:EPL2004} and van der Poel \textit{et al.}~\cite{Poel:JFM2013}.  

Our results are essentially numerical.  It will be useful to construct low-dimensional models of $\mathrm{Pr}=\infty$  convection, and study how the velocity in one of the perpendicular direction gets suppressed.  This work is under progress.

\section*{Acknowledgement} 
Our numerical simulations were performed at {\em hpc} and {\em newton} clusters of IIT Kanpur, and at {\em Param Yuva} cluster of CDAC Pune. This work was supported by a research grant SERB/F/3279/2013-14 from Science and Engineering Research Board, India. We thank Supriyo Paul for useful comments and sharing his earlier results on 2D RBC. We also thank A. Kumar for providing help in flow visualization.


\begin{thebibliography}{21}
\expandafter\ifx\csname natexlab\endcsname\relax\def\natexlab#1{#1}\fi
\expandafter\ifx\csname bibnamefont\endcsname\relax
  \def\bibnamefont#1{#1}\fi
\expandafter\ifx\csname bibfnamefont\endcsname\relax
  \def\bibfnamefont#1{#1}\fi
\expandafter\ifx\csname citenamefont\endcsname\relax
  \def\citenamefont#1{#1}\fi
\expandafter\ifx\csname url\endcsname\relax
  \def\url#1{\texttt{#1}}\fi
\expandafter\ifx\csname urlprefix\endcsname\relax\def\urlprefix{URL }\fi
\providecommand{\bibinfo}[2]{#2}
\providecommand{\eprint}[2][]{\url{#2}}

\bibitem[{\citenamefont{{Ahlers} et~al.}(2009)\citenamefont{{Ahlers},
  {Grossmann}, and {Lohse}}}]{Ahlers:RMP2009}
\bibinfo{author}{\bibfnamefont{G.}~\bibnamefont{{Ahlers}}},
  \bibinfo{author}{\bibfnamefont{S.}~\bibnamefont{{Grossmann}}},
  \bibnamefont{and} \bibinfo{author}{\bibfnamefont{D.}~\bibnamefont{{Lohse}}},
  \bibinfo{journal}{Rev. Mod. Phys.} \textbf{\bibinfo{volume}{81}},
  \bibinfo{pages}{503} (\bibinfo{year}{2009}).

\bibitem[{\citenamefont{{Schmalzl} et~al.}(2002)\citenamefont{{Schmalzl},
  {Breuer}, and {Hansen}}}]{Schmalzl:GAFD2002}
\bibinfo{author}{\bibfnamefont{J.}~\bibnamefont{{Schmalzl}}},
  \bibinfo{author}{\bibfnamefont{M.}~\bibnamefont{{Breuer}}}, \bibnamefont{and}
  \bibinfo{author}{\bibfnamefont{U.}~\bibnamefont{{Hansen}}},
  \bibinfo{journal}{Geophys. Astrophys. Fluid Dyn.}
  \textbf{\bibinfo{volume}{96}}, \bibinfo{pages}{381} (\bibinfo{year}{2002}).

\bibitem[{\citenamefont{{Schmalzl} et~al.}(2004)\citenamefont{{Schmalzl},
  {Breuer}, and {Hansen}}}]{Schmalzl:EPL2004}
\bibinfo{author}{\bibfnamefont{J.}~\bibnamefont{{Schmalzl}}},
  \bibinfo{author}{\bibfnamefont{M.}~\bibnamefont{{Breuer}}}, \bibnamefont{and}
  \bibinfo{author}{\bibfnamefont{U.}~\bibnamefont{{Hansen}}},
  \bibinfo{journal}{Europhys. Lett.} \textbf{\bibinfo{volume}{67}},
  \bibinfo{pages}{390} (\bibinfo{year}{2004}).

\bibitem[{\citenamefont{{van der Poel} et~al.}(2013)\citenamefont{{van der
  Poel}, {Stevens}, and {Lohse}}}]{Poel:JFM2013}
\bibinfo{author}{\bibfnamefont{E.~P.} \bibnamefont{{van der Poel}}},
  \bibinfo{author}{\bibfnamefont{R.~J. A.~M.} \bibnamefont{{Stevens}}},
  \bibnamefont{and} \bibinfo{author}{\bibfnamefont{D.}~\bibnamefont{{Lohse}}},
  \bibinfo{journal}{J. Fluid Mech.} \textbf{\bibinfo{volume}{736}},
  \bibinfo{pages}{177} (\bibinfo{year}{2013}).

\bibitem[{\citenamefont{{Lohse} and {Xia}}(2010)}]{Lohse:ARFM2010}
\bibinfo{author}{\bibfnamefont{D.}~\bibnamefont{{Lohse}}} \bibnamefont{and}
  \bibinfo{author}{\bibfnamefont{K.~Q.} \bibnamefont{{Xia}}},
  \bibinfo{journal}{Annu. Rev. Fluid Mech.} \textbf{\bibinfo{volume}{42}},
  \bibinfo{pages}{335} (\bibinfo{year}{2010}).

\bibitem[{\citenamefont{{Grossmann} and {Lohse}}(1992)}]{Grossmann:PRA1992}
\bibinfo{author}{\bibfnamefont{S.}~\bibnamefont{{Grossmann}}} \bibnamefont{and}
  \bibinfo{author}{\bibfnamefont{D.}~\bibnamefont{{Lohse}}},
  \bibinfo{journal}{Phys. Rev. A} \textbf{\bibinfo{volume}{46}},
  \bibinfo{pages}{903} (\bibinfo{year}{1992}).

\bibitem[{\citenamefont{{L'vov}}(1991)}]{Lvov:PRL1991}
\bibinfo{author}{\bibfnamefont{V.~S.} \bibnamefont{{L'vov}}},
  \bibinfo{journal}{Phys. Rev. Lett.} \textbf{\bibinfo{volume}{67}},
  \bibinfo{pages}{687} (\bibinfo{year}{1991}).

\bibitem[{\citenamefont{{L'vov} and {Falkovich}}(1992)}]{Lvov:PD1992}
\bibinfo{author}{\bibfnamefont{V.~S.} \bibnamefont{{L'vov}}} \bibnamefont{and}
  \bibinfo{author}{\bibfnamefont{G.}~\bibnamefont{{Falkovich}}},
  \bibinfo{journal}{Physica D} \textbf{\bibinfo{volume}{57}},
  \bibinfo{pages}{85} (\bibinfo{year}{1992}).

\bibitem[{\citenamefont{{Toh} and {Suzuki}}(1994)}]{Toh:PRL1994}
\bibinfo{author}{\bibfnamefont{S.}~\bibnamefont{{Toh}}} \bibnamefont{and}
  \bibinfo{author}{\bibfnamefont{E.}~\bibnamefont{{Suzuki}}},
  \bibinfo{journal}{Phys. Rev. Lett.} \textbf{\bibinfo{volume}{73}},
  \bibinfo{pages}{1501} (\bibinfo{year}{1994}).

\bibitem[{\citenamefont{{Vincent} and {Yuen}}(1999)}]{Vincent:PRE1999}
\bibinfo{author}{\bibfnamefont{A.~P.} \bibnamefont{{Vincent}}}
  \bibnamefont{and} \bibinfo{author}{\bibfnamefont{D.~A.}
  \bibnamefont{{Yuen}}}, \bibinfo{journal}{Phys. Rev. E}
  \textbf{\bibinfo{volume}{60}}, \bibinfo{pages}{2957} (\bibinfo{year}{1999}).

\bibitem[{\citenamefont{{Vincent} and {Yuen}}(2000)}]{Vincent:PRE2000}
\bibinfo{author}{\bibfnamefont{A.~P.} \bibnamefont{{Vincent}}}
  \bibnamefont{and} \bibinfo{author}{\bibfnamefont{D.~A.}
  \bibnamefont{{Yuen}}}, \bibinfo{journal}{Phys. Rev. E}
  \textbf{\bibinfo{volume}{61}}, \bibinfo{pages}{5241} (\bibinfo{year}{2000}).

\bibitem[{\citenamefont{{Mishra} and {Verma}}(2010)}]{Mishra:PRE2010}
\bibinfo{author}{\bibfnamefont{P.~K.} \bibnamefont{{Mishra}}} \bibnamefont{and}
  \bibinfo{author}{\bibfnamefont{M.~K.} \bibnamefont{{Verma}}},
  \bibinfo{journal}{Phys. Rev. E} \textbf{\bibinfo{volume}{81}},
  \bibinfo{pages}{056316} (\bibinfo{year}{2010}).

\bibitem[{\citenamefont{{Pandey} et~al.}(2014)\citenamefont{{Pandey}, {Verma},
  and {Mishra}}}]{Pandey:PRE2014}
\bibinfo{author}{\bibfnamefont{A.}~\bibnamefont{{Pandey}}},
  \bibinfo{author}{\bibfnamefont{M.~K.} \bibnamefont{{Verma}}},
  \bibnamefont{and} \bibinfo{author}{\bibfnamefont{P.~K.}
  \bibnamefont{{Mishra}}}, \bibinfo{journal}{Phys. Rev. E}
  \textbf{\bibinfo{volume}{89}}, \bibinfo{pages}{023006}
  (\bibinfo{year}{2014}).

\bibitem[{\citenamefont{{Silano} et~al.}(2010)\citenamefont{{Silano},
  {Sreenivasan}, and {Verzicco}}}]{Silano:JFM2010}
\bibinfo{author}{\bibfnamefont{G.}~\bibnamefont{{Silano}}},
  \bibinfo{author}{\bibfnamefont{K.~R.} \bibnamefont{{Sreenivasan}}},
  \bibnamefont{and}
  \bibinfo{author}{\bibfnamefont{R.}~\bibnamefont{{Verzicco}}},
  \bibinfo{journal}{J. Fluid Mech.} \textbf{\bibinfo{volume}{662}},
  \bibinfo{pages}{409} (\bibinfo{year}{2010}).

\bibitem[{\citenamefont{{Verma} et~al.}(2013)\citenamefont{{Verma},
  {Chatterjee}, {Reddy}, {Yadav}, {Paul}, {Chandra}, and
  {Samtaney}}}]{Verma:Pramana2013}
\bibinfo{author}{\bibfnamefont{M.~K.} \bibnamefont{{Verma}}},
  \bibinfo{author}{\bibfnamefont{A.~G.} \bibnamefont{{Chatterjee}}},
  \bibinfo{author}{\bibfnamefont{K.~S.} \bibnamefont{{Reddy}}},
  \bibinfo{author}{\bibfnamefont{R.~K.} \bibnamefont{{Yadav}}},
  \bibinfo{author}{\bibfnamefont{S.}~\bibnamefont{{Paul}}},
  \bibinfo{author}{\bibfnamefont{M.}~\bibnamefont{{Chandra}}},
  \bibnamefont{and}
  \bibinfo{author}{\bibfnamefont{R.}~\bibnamefont{{Samtaney}}},
  \bibinfo{journal}{Pramana} \textbf{\bibinfo{volume}{81}},
  \bibinfo{pages}{617} (\bibinfo{year}{2013}).

\bibitem[{\citenamefont{{Fischer}}(1997)}]{Fischer:JCP1997}
\bibinfo{author}{\bibfnamefont{P.~F.} \bibnamefont{{Fischer}}},
  \bibinfo{journal}{J. Comp. Phys.} \textbf{\bibinfo{volume}{133}},
  \bibinfo{pages}{84} (\bibinfo{year}{1997}).

\bibitem[{\citenamefont{{Chandra} and {Verma}}(2013)}]{Chandra:PRL2013}
\bibinfo{author}{\bibfnamefont{M.}~\bibnamefont{{Chandra}}} \bibnamefont{and}
  \bibinfo{author}{\bibfnamefont{M.~K.} \bibnamefont{{Verma}}},
  \bibinfo{journal}{Phys. Rev. Lett.} \textbf{\bibinfo{volume}{110}},
  \bibinfo{pages}{114503} (\bibinfo{year}{2013}).

\bibitem[{\citenamefont{{Vitanov}}(1998)}]{Vitanov:PLA1998}
\bibinfo{author}{\bibfnamefont{N.~K.} \bibnamefont{{Vitanov}}},
  \bibinfo{journal}{Phys. Lett. A} \textbf{\bibinfo{volume}{248}},
  \bibinfo{pages}{338} (\bibinfo{year}{1998}).

\bibitem[{\citenamefont{{Verma}}(2004)}]{Verma:PR2004}
\bibinfo{author}{\bibfnamefont{M.~K.} \bibnamefont{{Verma}}},
  \bibinfo{journal}{Phys. Rep.} \textbf{\bibinfo{volume}{401}},
  \bibinfo{pages}{229} (\bibinfo{year}{2004}).

\bibitem[{\citenamefont{{Boffetta} and {Ecke}}(2012)}]{Boffetta:ARFM2012}
\bibinfo{author}{\bibfnamefont{G.}~\bibnamefont{{Boffetta}}} \bibnamefont{and}
  \bibinfo{author}{\bibfnamefont{R.~E.} \bibnamefont{{Ecke}}},
  \bibinfo{journal}{Annu. Rev. Fluid Mech.} \textbf{\bibinfo{volume}{44}},
  \bibinfo{pages}{427} (\bibinfo{year}{2012}).

\bibitem[{\citenamefont{{Shraiman} and {Siggia}}(1990)}]{Shraiman:PRA1990}
\bibinfo{author}{\bibfnamefont{B.~I.} \bibnamefont{{Shraiman}}}
  \bibnamefont{and} \bibinfo{author}{\bibfnamefont{E.~D.}
  \bibnamefont{{Siggia}}}, \bibinfo{journal}{Phys. Rev. A}
  \textbf{\bibinfo{volume}{42}}, \bibinfo{pages}{3650} (\bibinfo{year}{1990}).

\end{thebibliography}
\end{document}